\documentclass[aps,prx,twocolumn,showpacs,psfig,superscriptaddress,longbibliography]{revtex4-1}
\usepackage{amssymb}
\usepackage{algpseudocode}
\usepackage{algorithm}
\usepackage{amsmath}
\usepackage{braket}
\usepackage{booktabs}
\usepackage{chngcntr}
\usepackage{color}
\usepackage{comment}
\usepackage{dsfont}
\usepackage{etoolbox}
\usepackage{epstopdf}
\usepackage[T1]{fontenc}
\usepackage{float}
\usepackage{graphicx}
\usepackage[colorlinks=true,linkcolor=blue,citecolor=blue,urlcolor=blue]{hyperref}
\usepackage[latin9]{inputenc}
\usepackage{indentfirst}
\usepackage{latexsym,bm,euscript}
\usepackage{mathrsfs}
\usepackage{multirow}
\usepackage{natbib}
\usepackage{soul}
\usepackage{subfigure}
\usepackage{times}
\usepackage{tikz}
\usepackage{txfonts}
\usepackage{ulem}
\usepackage{xspace}
\usepackage{xfrac}

\newcommand{\Sec}[1]{Sec.~\ref{#1}}

\newcommand{\Eq}[1]{Eq.~\eqref{#1}}

\newcommand{\Fig}[1]{Fig.~\ref{#1}}

\graphicspath{{./Figs/}}

\begin{document}
\title{Universal Thermodynamics in the Kitaev Fractional Liquid}

\author{Han Li}
\affiliation{School of Physics, Key Laboratory of Micro-Nano Measurement-Manipulation
and Physics (Ministry of Education), Beihang University, Beijing 100191, China}

\author{Dai-Wei Qu}
\affiliation{School of Physics, Key Laboratory of Micro-Nano Measurement-Manipulation
and Physics (Ministry of Education), Beihang University, Beijing 100191, China}

\author{Hao-Kai Zhang}
\affiliation{School of Physics, Key Laboratory of Micro-Nano Measurement-Manipulation
and Physics (Ministry of Education), Beihang University, Beijing 100191, China}

\author{Yi-Zhen Jia}
\affiliation{School of Physics, Key Laboratory of Micro-Nano Measurement-Manipulation
and Physics (Ministry of Education), Beihang University, Beijing 100191, China}


\author{Shou-Shu Gong}
\email{shoushu.gong@buaa.edu.cn}
\affiliation{School of Physics, Key Laboratory of Micro-Nano Measurement-Manipulation
and Physics (Ministry of Education), Beihang University, Beijing 100191, China}

\author{Yang Qi}
\email{qiyang@fudan.edu.cn}
\affiliation{State Key Laboratory of Surface Physics, Fudan University, Shanghai 200433, China}
\affiliation{Center for Field Theory and Particle Physics, Department of Physics, Fudan University, Shanghai 200433, China}
\affiliation{Collaborative Innovation Center of Advanced Microstructures, Nanjing 210093, China}

\author{Wei Li}
\email{w.li@buaa.edu.cn}
\affiliation{School of Physics, Key Laboratory of Micro-Nano Measurement-Manipulation
and Physics (Ministry of Education), Beihang University, Beijing 100191, China}
\affiliation{International Research Institute of Multidisciplinary Science, Beihang University, Beijing 100191, China}

\begin{abstract}
In the Kitaev honeycomb model, the quantum spin fractionalizes
into itinerant Majorana and gauge flux spontaneously upon cooling,
leading to rich experimental ramifications at finite temperature
and an upsurge of research interest.
In this work, we employ the exponential tensor renormalization group approach
to explore the Kitaev model under various perturbations,
including the external fields, Heisenberg, and the off-diagonal couplings
that are common in the Kitaev materials. Through large-scale manybody calculations,
we find a Kitaev fractional liquid at intermediate temperature that is robust against perturbations.
The fractional liquid exhibits universal thermodynamic behaviors,
including the fractional thermal entropy, metallic specific heat,
and an intermediate-temperature Curie law of magnetic susceptibility.
The emergent universal susceptibility behavior, with a modified Curie constant,
can be ascribed to the strongly fluctuating $\mathbb{Z}_2$ fluxes 
as well as the extremely short-ranged and bond-directional spin correlations.
With this insight, we revisit the susceptibility measurements of
Na$_2$IrO$_3$ and $\alpha$-RuCl$_3$, and find evident signatures
of finite-temperature fractionalization and ferromagnetic Kitaev couplings. 
Moreover, the peculiar spin correlation in the fractional liquid 
corresponds to a stripy structure factor which rotates in the extended Brillouin zone
as the spin component changes. Therefore, our findings encourage
future experimental exploration of fractional liquid in the Kitaev materials
by thermodynamic measurements and spin-resolved structure factor probes.
\end{abstract}

\date{\today}
\maketitle

\section{Introduction}
Quantum spin liquids realize a novel class of quantum states of matter \cite{Anderson1973,Balents2010,Zhou2017}
which represent the nonmagnetic states with emergent long-range entanglement
and fractionalized excitations in frustrated magnets \cite{Ramirez2003,Lacroix2011}.
After extensive search for decades, the exactly soluble Kitaev model on the honeycomb lattice \cite{Kitaev2006},
which has three types of bond-directional Ising couplings in the trivalent, was found as a prototypical
system that can realize both gapped and gapless spin liquid states with
spins fractionalized into itinerant and localized Majorana fermions \cite{Kitaev2006,Hermanns2018}.
In particular, in the gapless spin liquid phase, time-reversal-symmetry breaking perturbations such
as a magnetic field can immediately induce a gapped non-Abelian chiral spin liquid, which has
potential application for topological quantum computation \cite{Kitaev2003}.

\begin{figure}[h!]
\includegraphics[angle=0,width=0.95\linewidth]{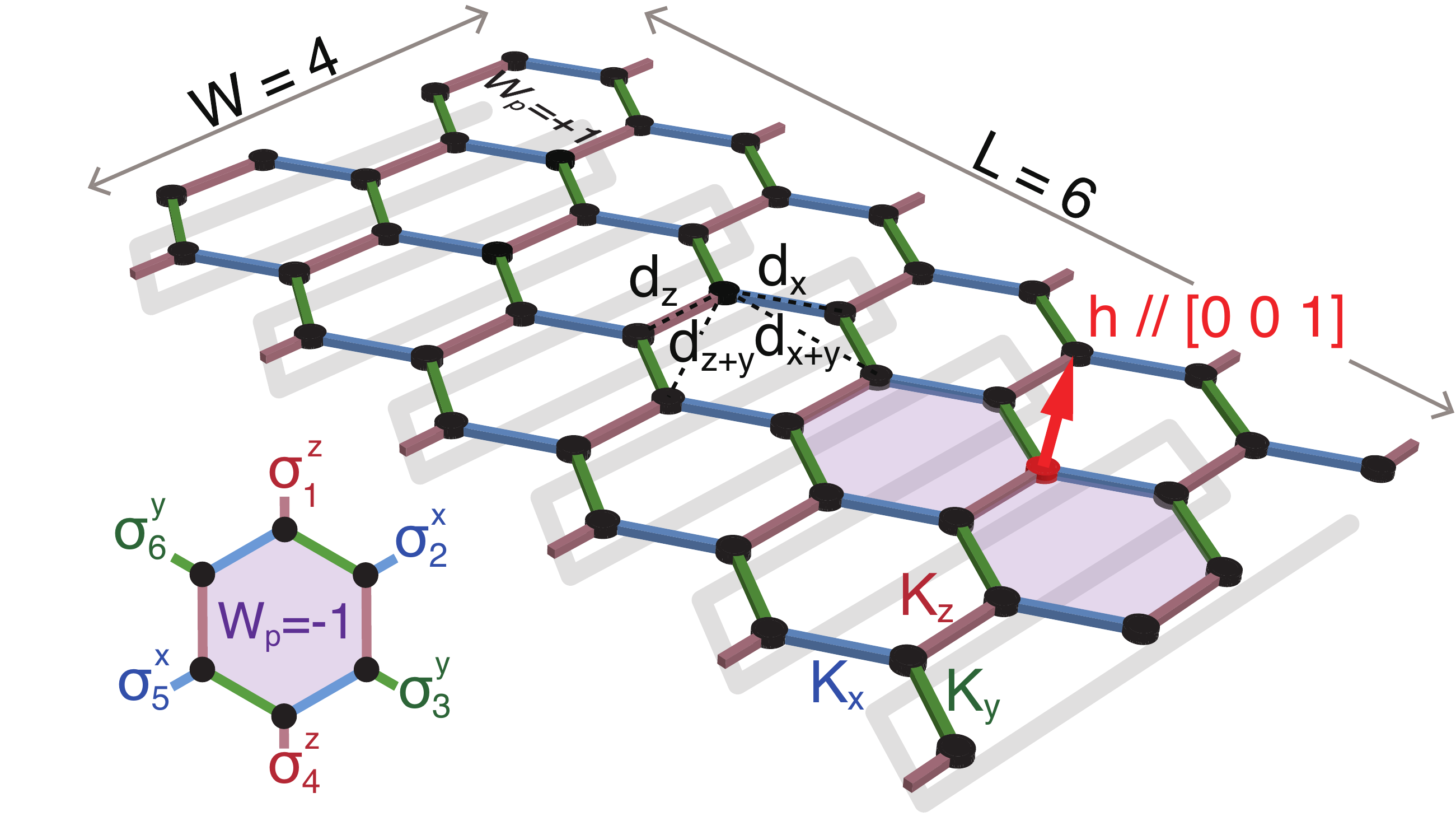}
\renewcommand{\figurename}{\textbf{Fig. }}
\caption{The cylindrical honeycomb lattice of size YC$4\times6\times2$
is illustrated with the snake-like quasi-1D mapping path (underlying grey
lines), adopted in the XTRG calculations. 
Three types of Kitaev bonds $K_x$, $K_y$ and $K_z$
are marked in blue, green, and red colors, respectively.
The $\mathbb{Z}_2$ flux
$W_P=\sigma_1^z  \sigma_2^x  \sigma_3^y  \sigma_4^z
 \sigma_5^x  \sigma_6^y = + 1$ (hexagon in white) or $= -1$ (purple),
with $\sigma_i^\gamma = 2 S_i^\gamma$,
can be pair-flipped by applying a magnetic field $h$
// [0 0 1] in spin-space cubic axis. Typical nearest and
next-nearest bonds, $d_x$, $d_z$, $d_{x+y}$,
and $d_{x+z}$, are indicated by the dashed lines.
}
\label{FigIllus}
\end{figure}

To pursue the materialization of the intriguing Kitaev spin liquid (KSL) states,
a race has been triggered for searching crystalline Kitaev-like materials
\cite{Trebst2017arXiv,Winter2017,Takagi2019}.
Fortunately, there are Mott insulators with strong spin-orbit coupling
and effective pseudospin $J_{\rm eff} = 1/2$ on the honeycomb lattice
that can realize the highly anisotropic couplings
and have been intensively studied in recent years, including the
iridate family such as A$_2$IrO$_3$ (A = Li and Na)
\cite{Jackeli2009,Chaloupka2010,Singh2010,Singh2012,Chaloupka2013,Kim2015,Mehlawat2017}
and H$_3$LiIr$_2$O$_6$ \cite{Kitagawa2018}, as well as the
ruthenate family such as $\alpha$-RuCl$_3$ \cite{Plumb2014,Sears2015,Banerjee2016,Banerjee2017,Do2017}.
Nonetheless, due to the couplings beyond the Kitaev model such as the Heisenberg, off-diagonal Gamma,
and further-neighbor interactions in real materials, semi-classical magnetic orders
appear at low temperature ($T$) in most of these KSL candidate compounds.
Surprisingly, signatures of fractionalization in quantum spins can nevertheless be detected
at intermediate temperature in some Kitaev materials
\cite{Kim2015,Banerjee2016,Banerjee2017,Do2017},
even when the low-$T$ states are classically ordered.
These exciting observations and the spin liquid phase found in the presence of magnetic fields
strongly suggest that these materials should
be in proximity to the KSL phase, and imminently call for
systematic understanding on the thermodynamic
properties of these Kitaev materials.

For the pure Kitaev model, the finite-$T$ fractionalization of spins
has been well-established via the Majorana-based quantum Monte Carlo (QMC) simulations
\cite{Nasu2014,Nasu2015,Motome2020}, which found two temperature scales
and an intermediate regime with unconventional properties
\cite{Nasu2014,Nasu2015,Yoshitake2016, Yoshitake2017b,Motome2020}.
This may shed light on the understanding of the exotic equilibrium \cite{Do2017}
and transport measurements \cite{Kasahara2018} in the Kitaev materials.
Nonetheless, the gap between experiments and theoretical understanding is still apparent.
Since the non-Kitaev interactions and magnetic field \cite{Winter2018,Janssen2019} spoil the
exact solution and restrict QMC simulations within relatively high temperature \cite{Motome2019},
little is known on the fractionalization of spins in the perturbed Kitaev models.

In this work, we employ the exponential tensor renormalization group (XTRG) \cite{Chen2018,Lih2019}
at finite temperature to explore the Kitaev model with various additional perturbations.
We find that the well-recognized characteristic features at intermediate temperature,
including the fractional thermal-entropy plateau and the linear-$T$
specific heat scaling clearly persist. More interestingly,
we find an emergent universal Curie law of magnetic susceptibility
below a high-$T$ scale $T_H$ (and above the low-$T$ scale $T_L$),
due to the fluctuating gauge fluxes and the predominant bond-directional
nearest-neighboring (NN) spin correlations.
Through exploring these robust finite-$T$ features,
we find a generic scenario in the perturbed Kitaev model:
A fractionalization of the quantum spin takes place spontaneously
as temperature decreases, and half of the spin degrees of freedom,
i.e. the itinerant Majorana fermions,
release large part of the entropy below $T_H$,
accompanied by the saturated short-range spin correlation;
meanwhile the other half, the gauge fluxes,
fluctuate strongly until near the much lower temperature $T_L$.
In between the two temperature scales, there reside a unique finite-$T$
quantum state dubbed the \textit{Kitaev fractional liquid}.

Based on our systematic characterization, we reveal that the fractional liquid regime does
exist at intermediate temperature, for an extended range of external fields or non-Kitaev terms
including the Heisenberg or Gamma couplings, even when the KSL ground state
is altered. These results therefore pave the way towards filling the gap
between theoretical and experimental studies of the Kitaev physics.
In particular, with insight from the universal susceptibility scaling,
we revisit the magnetic susceptibility data of Na$_2$IrO$_3$ and $\alpha$-RuCl$_3$
and find evident signature of Curie susceptibility scaling
in the two prominent Kitaev materials, suggesting the existence of the fractional liquid therein.
We also propose the spin-correlation diagnostics of the fractional quantum liquid,
which is available through spin-resolved neutron or X-ray diffusion scatterings.

The rest part of the paper is organized as follows.
In \Sec{SecModMeth}, we introduce the Kitaev model with
perturbative couplings and the XTRG method for finite-$T$
simulations. Secs. \ref{SecPKFL}, \ref{SecSus}, and \ref{SecKFLpert} present our main results on the
universal thermodynamic properties and spin-correlation
characterization of the intermediate-$T$ Kitaev fractional liquid that is
robust against additional interactions. In \Sec{SecSusKMater},
we discuss the relevance of universal susceptibility scaling
in two Kitaev materials, Na$_2$IrO$_3$ and $\alpha$-RuCl$_3$,
and propose the spin structure diagnosis of the fractional liquid.
Finally, we summarize in \Sec{SecConOut} our main conclusions
and discuss their implications in future studies
of the Kitaev model and materials.

\section{Model and methods}
\label{SecModMeth}

We study the celebrated Kitaev honeycomb model with various perturbative terms as follows
\begin{equation}
H = \sum_{\langle i,j \rangle _{\gamma}} [K_{\gamma} S_i^{\gamma} S_j^{\gamma} +
J \, \textbf{S}_i \cdot \textbf{S}_j + \Gamma(S_i^{\alpha} S_j^{\beta} + S_i^{\beta} S_j^{\alpha})] +
\sum_i h \ S_i^z,
\label{EqHamil}
\end{equation}
where $\textbf{S}_i = \{ S^x_i, S^y_i, S^z_i \}$ represent the spin-$1/2$ vector operator at site $i$.
The three cubic coordinates $\{ \alpha, \beta, \gamma \}$ represent $\{ x,y,z \}$ up to a cyclic permutation,
and $\langle i , j\rangle_{\gamma}$ denotes the NN $\gamma$-bond [see Fig.~\ref{FigIllus}]
with an Ising coupling $K_{\gamma}$. 
The non-Kitaev perturbations include the NN isotropic Heisenberg coupling $J$
and an off-diagonal Gamma ($\Gamma$) interaction.
Moreover, we apply an external field $h$ coupled to the $S^z$ component,
which is along [0 0 1] spin-space direction and tilts an angle of 45$^{\circ}$
from the honeycomb ${ab}$ plane (see Fig.~\ref{FigIllus})
in typical Kitaev materials.

For the pure Kitaev model, spin can be represented as localized ($b^\gamma_i$)
and itinerant Majorana fermions ($c_i$) through the parton construction
$S_i^\gamma = \frac{i}{2} b_i^\gamma c_i$ \cite{Kitaev2003}.
The itinerant Majorana can be related to the building up of spin correlation
$\langle S_i^z S_j^z \rangle = u_{ij}^z \langle c_i c_j \rangle$, and
$u_{ij}^\gamma = b_i^\gamma b_j^\gamma = \pm i$ represents
a $\mathbb{Z}_2$ gauge connection.
The product of $u_{ij}^\gamma$ on the six edges of a hexagon gives rise to
the flux $W_P = \pm 1$ as illustrated in \Fig{FigIllus},
which conserves in the pure Kitaev case.

For the typical Kitaev materials Na$_2$IrO$_3$ and $\alpha$-RuCl$_3$,
the predominant Kitaev interactions have been widely accepted
to be ferromagnetic (FM) \cite{Winter2016,Do2017,Winter2017,Winter2017b,Motome2019,Sears2020}.
Thus, we choose $K_{\gamma} < 0$ henceforth and set $K_z = -1$ as the fixed energy scale.
The Gamma term is suggested to be of next-to-leading order
and with $\Gamma > 0$ in $\alpha$-RuCl$_3$ \cite{Rau2014,Ran2017}.
On the other hand, in Na$_2$IrO$_3$ the Heisenberg coupling is
believed to play an important role and is proposed to be antiferromagnetic (AF) \cite{Katukuri2014,Winter2016}.
Thus, in our simulations below we introduce additional AF 
Heisenberg or Gamma term with $J,\Gamma > 0$, and focus on the finite-$T$ quantum
states of the perturbed Kitaev model.

\begin{figure}[h!]
\includegraphics[angle=0,width=1\linewidth]{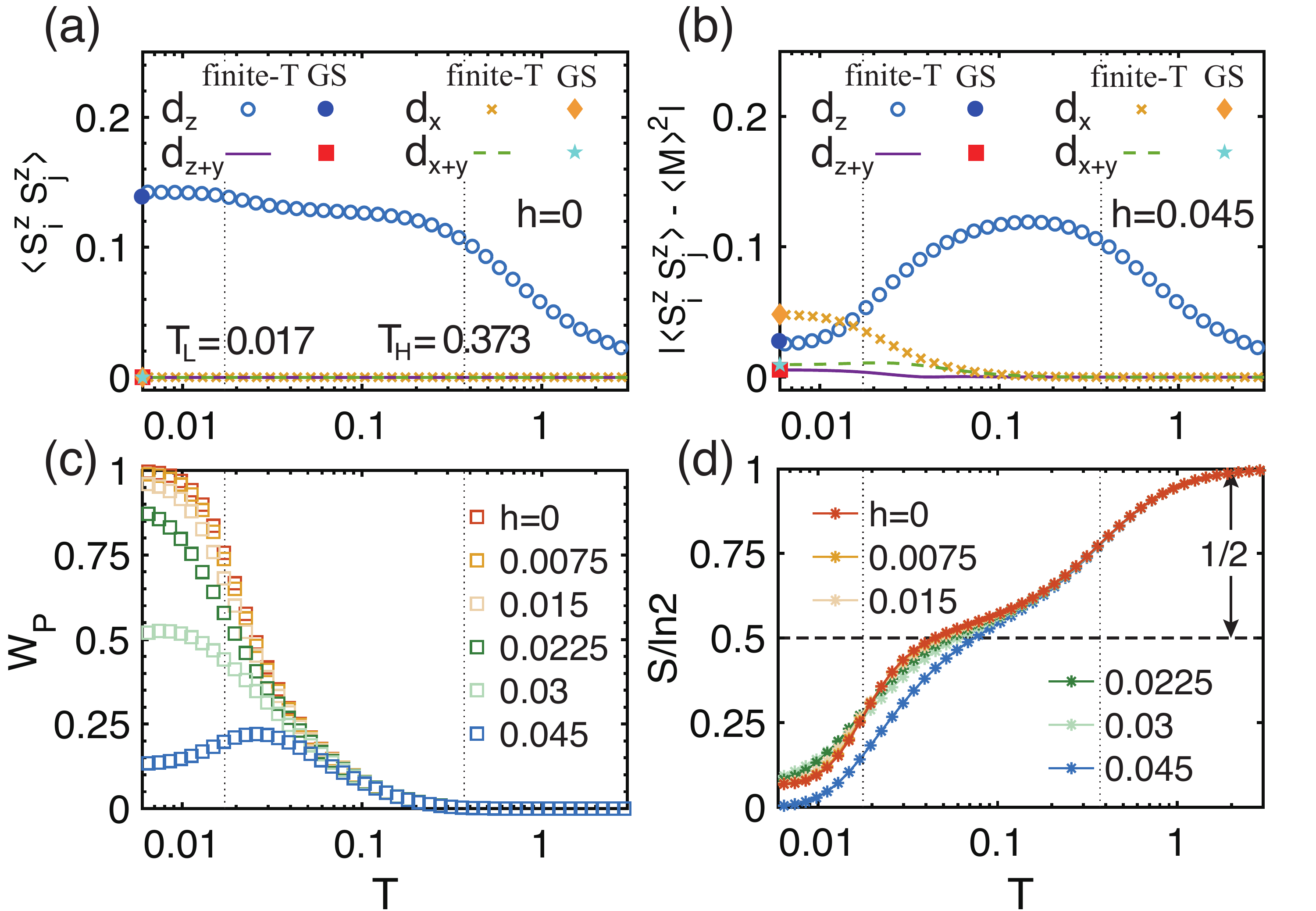}
\renewcommand{\figurename}{\textbf{Fig. }}
\caption{The spin correlations $\langle S_i^z S_j^z \rangle$
with (a) $h=0$ and (b) $0.045$ on bond $\langle i,j \rangle = d_x$, $d_z$, $d_{x+y}$,
and $d_{x+z}$ as depicted in \Fig{FigIllus},
where the NN spin correlations
are established at $T_H \simeq 0.373$, and converge to the
ground-state (GS) DMRG results (solid symbols) at low temperature.
In panel (b), the mean magnetization squared
$\langle M \rangle^2$ has been subtracted in the correlation data (see main text).
(c) Thermal average of flux $W_{\rm P}$ and (d)
thermal entropy $S$ results under various fields.
As temperature decreases, the flux increases monotonically for $h\lesssim h_c$
and builds up rapidly at around $T_L \simeq 0.017$, while it becomes
non-monotonic for $h\gtrsim h_c$ and starts to decreases below
a certain low temperature after the initial stage of increasing.
(d) The entropy with $h\lesssim h_c$
exists a stable $S\simeq\frac{1}{2} \ln{2}$ quasi-plateau,
above which the curves almost coincide;
while for $h\gtrsim h_c$, the entropy leaves a shoulder-like structure
at intermediate temperature.
}
\label{FigCurves}
\end{figure}

In the XTRG simulations \cite{Chen2018,Lih2019},
we follow the snake-like path and
map the cylinder into a quasi-one-dimensional (1D) system
as illustrated by the grey lines in the shadow of Fig.~\ref{FigIllus}.
We work on the cylindrical lattice YC\,$W\times L\times 2$
(see \Fig{FigIllus}), with width $W$ (up to 4) and length $L$ (up to 6),
thus total site $N=2WL$, which reproduces results
in excellent agreement with benchmark exact diagonalization (ED)
on small system size as well as large-scale Majorana
QMC simulations (see Appendix~\ref{App:XTRG}).
From comprehensive benchmark results,
we see finite-size effects are in most cases modest,
at least for the intermediate-$T$ regime of main focus in the current study.
This is in contrast from the ground-state calculations and
can be ascribed to the extremely short-range correlations
at intermediate temperature, even under perturbations
(see Appendix~\ref{App:CorrFunc}).

Moreover, as there are two characteristic temperature
scales in the Kitaev model that are typically separated
by more than one orders of magnitude,
our XTRG approach that cools down the system
exponentially in temperature scale
turns out to be very efficient to address this problem.
Therefore, we can exploit the XTRG method
that is free of sign problem in simulating the Kitaev model
under various perturbations  down to extremely low temperature, $T \simeq 0.005$.
In practical calculations we retain bond dimension up to $D=600$
that is capable to describe the thermal quantum states accurately and
produce well converged thermodynamic results (Appendix~\ref{AppLogEnt}).
We have also employed the density matrix renormalization group (DMRG) method
to simulate the ground state, confirming the XTRG results at very low temperature.

\begin{figure}[t!]
\includegraphics[angle=0,width=1\linewidth]{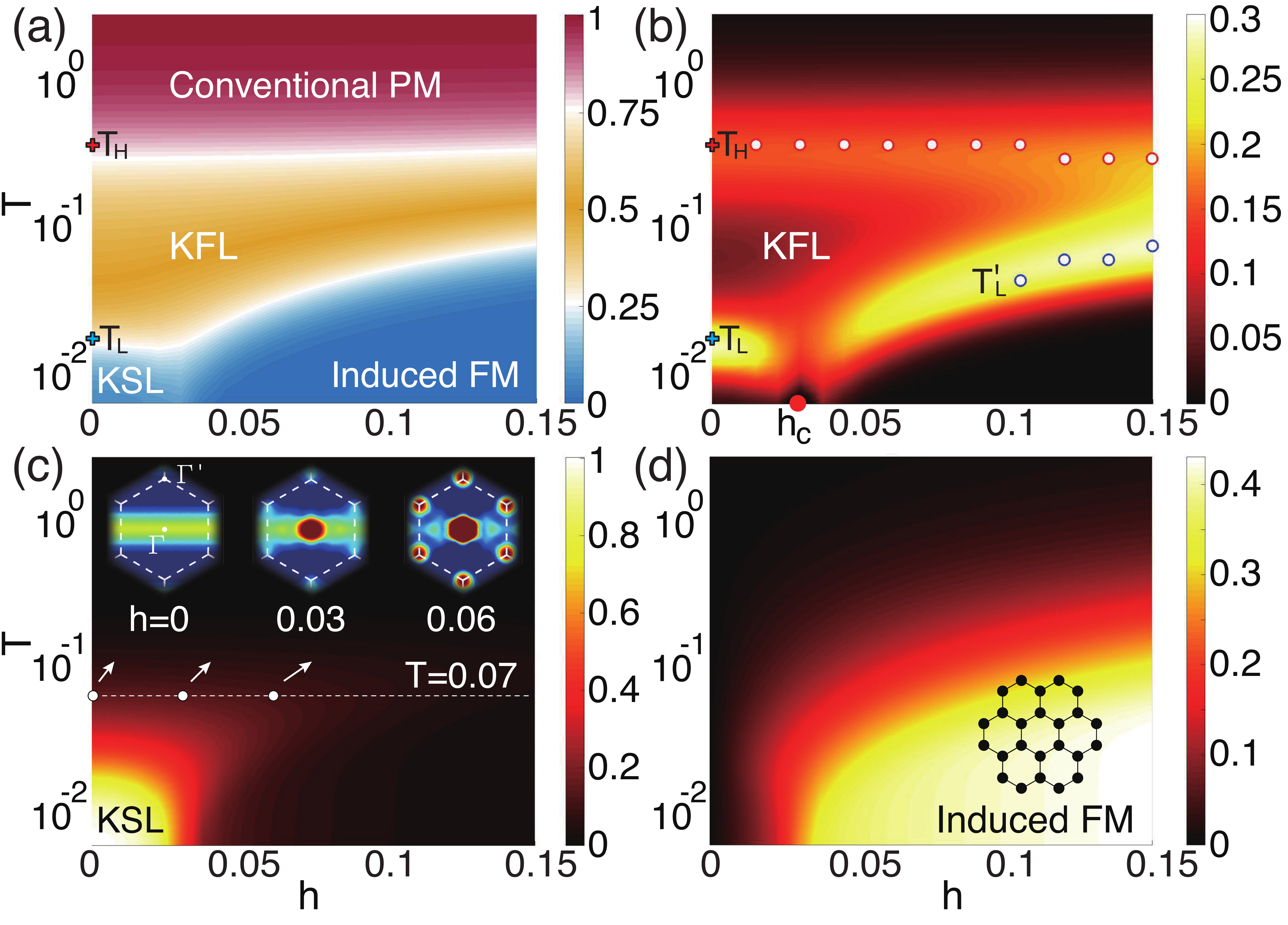}
\renewcommand{\figurename}{\textbf{Fig. }}
\caption{The contour plot of (a) thermal entropy $S/\ln{2}$ and (b) specific heat of the Kitaev model
under magnetic fields $h$, where two crossover temperature scales and
an intermediate Kitaev fractional liquid (KFL) regime are evident.
The $T_H$ and $T_L$ ($T_L'$) determined from (b)
coincide with results from Ref.~\cite{Motome2019} (the white dots,
available for $T \gtrsim 0.05$ for $h\neq0$).
(c) The flux $W_{\rm P}$ and
(d) magnetization $M$ are shown vs. various $h$ and $T$.
Three finite-$T$ ($T=0.07$)
spin structure factors at $h=0$, $0.03$ and $0.06$ are presented in the inset of (c)
(whose intensity color bar ranges from 0.2 to 0.5).
A structure peak emerges at center ($\Gamma$ point) of the extended Brillouin zone
due to the induced FM moment $M$ at $h>0$, together with a stripy background.
(d) clearly indicates the field-induced FM phase,
and the black dots represent the spins parallel to the [0 0 1] field in the inset.
}
\label{FigPhsDigFld}
\end{figure}

\section{Kitaev fractional liquid}
\label{SecPKFL}
\subsection{Ferromagnetic Kitaev model at finite temperature}
It has been well-established that the itinerant and localized Majorana fermions
can give rise to the two temperature scales, $T_H$ and $T_L$, respectively.
At intermediate temperature $T_L \lesssim T \lesssim T_H$,
an exotic fractional quantum liquid emerges.
To start with, we simulate finite-$T$ properties of the isotropic FM
Kitaev model ($K_{x,y,z}=-1$) by XTRG on a YC$4\times6\times2$ cylinder.
In \Fig{FigCurves}(a), we plot the spin correlation
$\langle S_i^z S_j^z \rangle$ measured
along the NN bonds $d_z, d_x$
($d_y$ always leads to the same results as $d_x$, thus not shown)
and next-nearest ones $d_{z+y}$ and $d_{x+y}$
(as depicted in \Fig{FigIllus}). From the zero-field XTRG results,
we find only the nearest $d_z$-bond correlations is established
below $T_H$, and the rest remain small down to the lowest temperature.
Thus the spin correlations $\langle S_i^\gamma S_j^\gamma \rangle$
are nonzero only on the NN $\gamma$-bond,
i.e., they are extremely short-range and bond-oriented,
in full agreement with the analytical results \cite{Baskaran2007, Feng2007}.

Matter of fact, the itinerant Majorana fermions
release most of their entropy and the NN
spin correlation is established at around $T_H$ [\Fig{FigCurves}(a)].
On the other hand, the rest ``half'' of spins, $\mathbb{Z}_2$ fluxes,
strongly fluctuate until the low-$T$ scale $T_L$,
as evidenced by small $W_{\rm P}$ above $T_L$  in \Fig{FigCurves}(c).
Correspondingly, in \Fig{FigCurves}(d)
the thermal entropy exhibits a quasi-plateau with $S\simeq \frac{1}{2} \ln{2}$,
which constitutes a thermodynamic signature of
the intermediate-$T$ Kitaev fractional liquid.

\subsection{Kitaev fractional liquid at finite fields}
Recently, the magnetic field effects on the Kitaev materials have raised great interest
\cite{Zheng2017,Winter2018,Kasahara2018,Janssen2019,Motome2019}.
For example, thermal quantum Hall signature has been detected
in the field-induced spin liquid phase of $\alpha$-RuCl$_3$ \cite{Kasahara2018}.
Given the fractional liquid scenario in pure Kitaev model,
it is also interesting to further study how it evolves under external fields \cite{Motome2019}.
In the following, we mainly focus on the field direction $h$ // [0 0 1],
and the effects of external fields along other directions are
discussed in Appendix~\ref{AppFieldOrients}, where qualitatively the same results are seen.

\begin{figure}[t!]
\includegraphics[angle=0,width=1\linewidth]{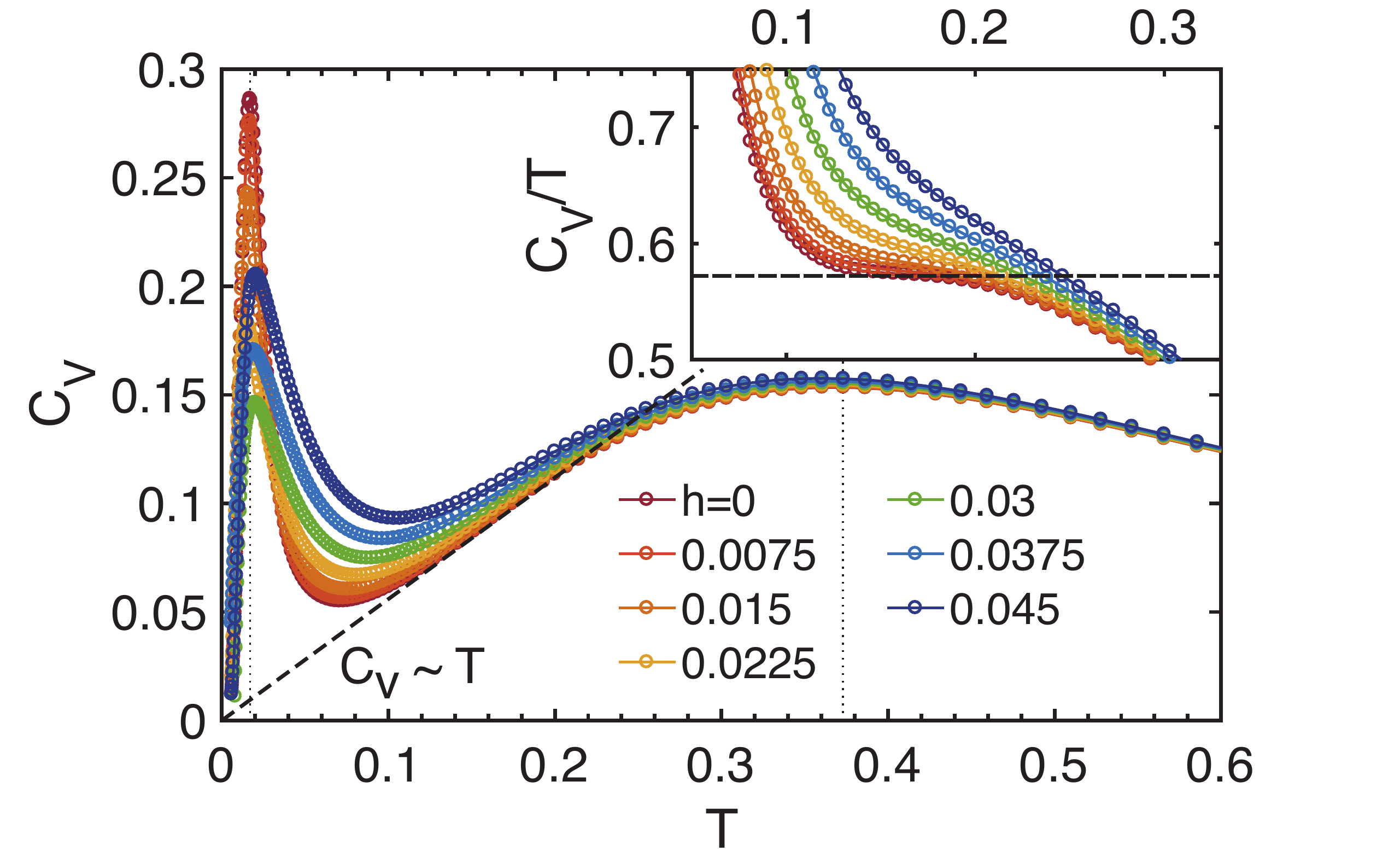}
\renewcommand{\figurename}{\textbf{Fig. }}
\caption{Specific heat $C_V$ under various magnetic fields $h$.
The linear-$T$ dependence of $C_V$ at intermediate temperature
is evident as indicated by the black dashed line,
which also underlines the quasi-plateau in the $C_V / T$
curves plotted in the inset.}
\label{FigCv}
\end{figure}

The finite-field entropy results are contour plotted in \Fig{FigPhsDigFld}(a),
with the $S$ vs. $T$ curves shown explicitly in \Fig{FigCurves}(d).
The two isentropic lines with $S/\ln2 = 0.75$ and $0.25$ in \Fig{FigPhsDigFld}(a)
correspond to the crossover temperatures
$T_H$ and $T_L$ (or $T'_L$) very well.
The intermediate-$T$ regime, which corresponds to the $\frac{1}{2}\ln2$
entropy plateau and signifies the fractional liquid,
extends to a rather wide field range regardless if
the low-$T$ phase is an asymptotic KSL or not.
The two-step release of entropy gives rise to the double-peaked
specific heat curves as illustrated in Figs. \ref{FigPhsDigFld}(b)
and \ref{FigCv}.
 
While $T_H$ remains stable as the magnetic field $h$ increases,
the low-$T$ crossover temperature $T_L$ drops
at $h = h_c \simeq 0.02$, and rises up again as $h$
further increases [$T_L'$ in \Fig{FigPhsDigFld}(b)].
As a matter of fact, $T'_L$ represents a different low-$T$ scale
since for $h>h_c$ the ground state is no longer KSL but a topologically
trivial field-induced FM phase. This can be seen evidently in \Fig{FigPhsDigFld}(c,d),
where the KSL is maintained for $h\lesssim h_c$
[$W_P \neq 0$, see also \Fig{FigCurves}(c)],
while the field-induced moment $M = (\frac{1}{N} \sum_i \langle S_i^z \rangle) \neq 0$
becomes significant only for $h\gtrsim h_c$.
Thus the transition field $h_c \simeq 0.02$
is in very good agreement with previous
studies \cite{Jiang2011,Gohlke2018},
although [111] fields are usually considered.

\subsection{Spin correlation and structure factor}
In robust fractional liquid regime, we again observe
extremely short-ranged and {bond-directional spin correlations}
that characterize this exotic finite-$T$ quantum state 
even under external fields. In \Fig{FigCurves}(b),
we plot the spin correlations $\langle S_i^z S_j^z \rangle - M^2$ with
the field-induced uniform background subtracted.
Although the longer-range correlations rise up at lower temperature and spoil the
extremely short correlation length ``inherited'' from the pure Kitaev model,
we nevertheless can observe that the $d_z$ correlation is predominate
over a very wide range of intermediate temperatures.

Correspondingly, we compute the static spin structure factor from the equal-time correlation function
\begin{equation}
\label{EqSq}
S^\gamma(q) = \sum_j \langle S_{i_0}^\gamma S_j^\gamma \rangle_\beta \, e^{i q (r_j-r_0)},
\end{equation}
where $r_0$ is the position vector of fixed reference site ($i_0$) and $j$ 
(and thus vector $r_j$) runs over the lattice sites.
We observe in \Fig{FigPhsDigFld}(c) a stripy structure factor at intermediate temperature
for both $h=0$ and $h\neq0$, resembling that in the KSL ground state \cite{Kim2015,Gotfryd2017,Juraj2019}.
For $h > 0$, there exists an additional bright spot at the $\Gamma$ point, which represents
the field-induced FM background in the system and can be observed clearly at intermediate temperature.
Besides the two-point correlation, from Figs.~\ref{FigCurves}(c) and \ref{FigPhsDigFld}
we observe that the $\mathbb{Z}_2$ flux operator $W_{\rm P}$
remains small at intermediate temperature, suggesting that the gauge fluxes can still flip
``freely'' under various fields in the fractional liquid regime.

\subsection{Linear-$T$ specific heat in the itinerant Majorana metal}
In the Kitaev fractional liquid, itinerant Majorana couples to the
thermally fluctuating fluxes and leads to a finite density
of states emerging at the Fermi surface, which can be characterized by
a linear $T$-dependence behavior of the specific heat $C_V$ \cite{Nasu2015}.
It is therefore of great interest to check if this intermediate-$T$ Majorana metal state
could persist at finite external fields, by examining the specific heat behaviors.

In \Fig{FigCv}, we show the XTRG results of $C_V$ that
indeed exhibit linear behaviors in the lower part of the intermediate-$T$ regime,
in excellent agreement to previous QMC result at zero field \cite{Nasu2015},
and thus extends the conclusion there to a finite range of fields.
As temperature further decreases, for $h\lesssim h_c$
the $C_V$ curve deflects the linear-$T$ behavior slightly above $T_L$ (or $T'_L$),
and when the fields $h\gtrsim h_c$, the low-$T$ scale $T_L'$ rises up as $h$ increases,
resulting in the reducing of the temperature range
showing metallic specific heat. We reveal this behavior
more clearly in the inset of \Fig{FigCv}, where a plateau
appears in $C_V/T$ for $h$ up to $h_c\simeq0.02$,
and it then turns into a shoulder-like structure for larger fields up to $h = 0.045$,
a reminiscent of the itinerant Majorana metal.

\begin{figure}[t!]
\includegraphics[angle=0,width=0.9\linewidth]{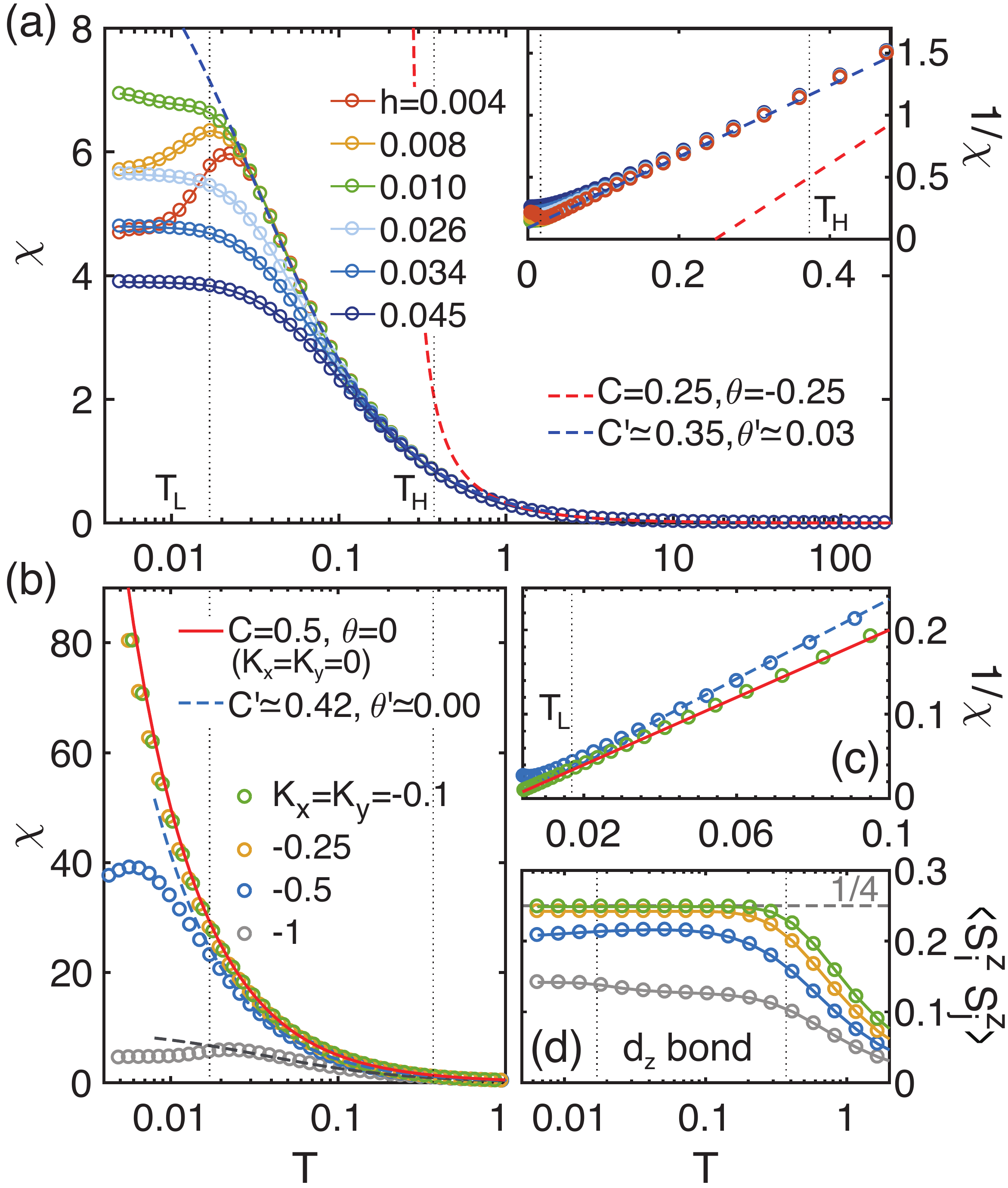}
\renewcommand{\figurename}{\textbf{Fig. }}
\caption{(a) Uniform magnetic susceptibility $\chi$ under various fields $h$,
with $1/\chi$ plotted in the inset.
The Curie-Weiss fittings in the high- and intermediate-$T$
regimes are indicated
by the red and blue dashed lines, respectively.
The high-$T$ behavior can be fitted
well by $1/(4T-J)$, and
immediately below $T\simeq T_H$ there emerges
a second Curie behavior.
(b) The zero-field susceptibility of the anisotropic model
with $K_x=K_y \neq K_z$,
where the emergent Curie behaviors are
observed in a wider temperature range.
The corresponding (c) $1/\chi$ fittings and
(d) spin correlations on the $d_z$ bond are also presented.
}
\label{FigSus}
\end{figure}

\section{Emergent Curie law in the fractional liquid}
\label{SecSus}

The linear-$T$ specific heat scaling constitutes an exotic thermodynamics
of the Kitaev fractional liquid. However, both itinerant Majorana and $\mathbb{Z}_2$ flux
have contributions in the specific heat and thus the Majorana metallic behavior is masked
by the entropy release of flux freezing at low temperature near $T_L$.
Moreover in experiments, there are contributions other than the Majorana fermions,
e.g., magnon excitations on top of the long-range magnetic order
and the excess phonon contribution not fully deducted,
that may enter the intermediate-$T$ specific heat data
and thus mask the expected universal linear-$T$ behavior.

Therefore, it calls for in-depth analysis of the other readily accessible thermodynamic
measurement in experiments that reflects more of the intrinsic
magnetic properties of the system, i.e., the susceptibility $\chi$.
From the analysis, we reveal that there emerges an universal Curie-law
behavior in the intermediate-$T$ susceptibility that is robust against perturbations
and origins from the fractionalization of spins.

\subsection{Magnetic susceptibility in the isotropic and anisotropic Kitaev models}

As a direct characterization of unconventional paramagnetism of the fractional liquid,
in \Fig{FigSus}(a), we show the susceptibility of the isotropic FM
Kitaev model under various magnetic fields.
For small fields $h \lesssim h_c$, the susceptibility data collapse into a
universal Curie form $\chi = \frac{C'}{T}$ in the fractional liquid regime,
further consolidated by plotting $1/\chi$ vs. $T$ in the inset of \Fig{FigSus}(a).
This unexpected universal Curie behavior extends across the fractional liquid regime,
spanning a very wide temperature range of about one order of magnitude ($T\sim0.03$-0.3),
constitutes an emergent phenomena distinct from conventional paramagnetism
at high temperature. The latter corresponds to a Curie constant
$C =S(S+1)/3=1/4$ relating to the intrinsic moment of spin $S=1/2$,
while the emergent Curie constant $C' \approx 1/3$ [see \Fig{FigSus}(a)]
in the fractional liquid regime, suggesting the presence of spin
correlations that renormalizes the effective magnetic moments.

Besides the isotropic case, we further turn to the anisotropic
Kitaev model and also find emergent Curie behaviors in \Fig{FigSus}(b,c)
that can extend down to even lower temperatures.
The Curie-Weiss fittings give $C' \approx 0.50$ for $K_{x,y} =-0.1$ and $-0.25$,
whose susceptibility is already very close to the ensemble
of isolated FM spin pairs [red lines in \Fig{FigSus}(b, c)],
and the Curie constant is close to twice a single spin-1/2, i.e., $C = 1/2$.
The emergent Curie susceptibility there can extend down to  
temperatures significantly lower than $T_L$ (of the isotropic case),
due to the reduction of flux gap. 
Therefore, starting from the trivial $K_{x,y}=0$ limit,
as we turn on the FM coupling $K_{x,y}$ in the system,
the Curie behavior remains, although the temperature range gets narrowed
and the Curie constant becomes modified.
For example, with $K_{x,y} =-0.5$, there also exists a Curie-law
regime with $C'$ reduced to $0.42$, as indicated by
the blue dashed lines in \Fig{FigSus}(b, c).
This universal Curie scaling extends all the way to the isotropic point,
implying spins can be pair-flipped nearly freely in the intermediate-$T$ regime.

Matter of fact, the emergent Curie behaviors can be observed as long as
the spin correlations $\langle S_i^z S_j^z \rangle$ are mostly restricted
within the $d_z$-bond [see \Fig{FigCurves}(a,b)] and in a quasi-plateau vs. $T$
at intermediate temperature shown in \Fig{FigSus}(d).
We note that, according to the perturbation theory,
the $\mathbb Z_2$ fluxes remain approximately conserved,
and spin correlations beyond the $d_z$-bonds only appear at the $h_c^2$ order.
At the same time, $\mathbb{Z}_2$ fluxes should be flippable in a
nearly free fashion, which can be observed as $W_P \ll 1$ in \Fig{FigCurves}(c).
These conditions are satisfied even when perturbative external fields are applied,
that gives rise to the intermediate-$T$ Curie law and will be elaborated in \Sec{SubSecFF} below.

\subsection{Fluctuating fluxes and diverging susceptibility}
\label{SubSecFF}
The diverging susceptibility reflects the strongly fluctuating fluxes
in the fractional liquid below $T_H$, and deviation occurs when the flux
starts to freeze out at about $T_L$.
To see this, we exploit the Kubo formula of susceptibility and assume
$M=0$ in the fractional liquid for the sake of discussion (Appendix~\ref{AppSusKitaev}),
\begin{equation}
\chi = \sum_{j} \int_{0}^{\beta}\langle S_{i_0}^z(\tau) \, S_j^z\rangle_{\beta}d\tau,
\label{EqDynaSus}
\end{equation}
where $S_{i_0}^z(\tau) = e^{\tau H} S_{i_0}^z e^{-\tau H}$,
and $j$ runs over NN sites of
$i_0$ by $d_z$ bond (as well as $i_0$ itself) in the fractional liquid regime
due to the extremely short-range correlations.
As illustrated in \Fig{FigIllus},
the spin operator $S_j^z$ flips a pair of fluxes on two neighboring
hexagons (color in purple) and changes the flux configuration of energy eigenstate.
After evolving along the imaginary time of $\tau$,
$S_{i_0}^z$ can flip the $\mathbb{Z}_2$ fluxes back and thus restore the
flux configurations, which results in a nonzero correlation.

\begin{figure*}[t!]
\includegraphics[angle=0,width=1\linewidth]{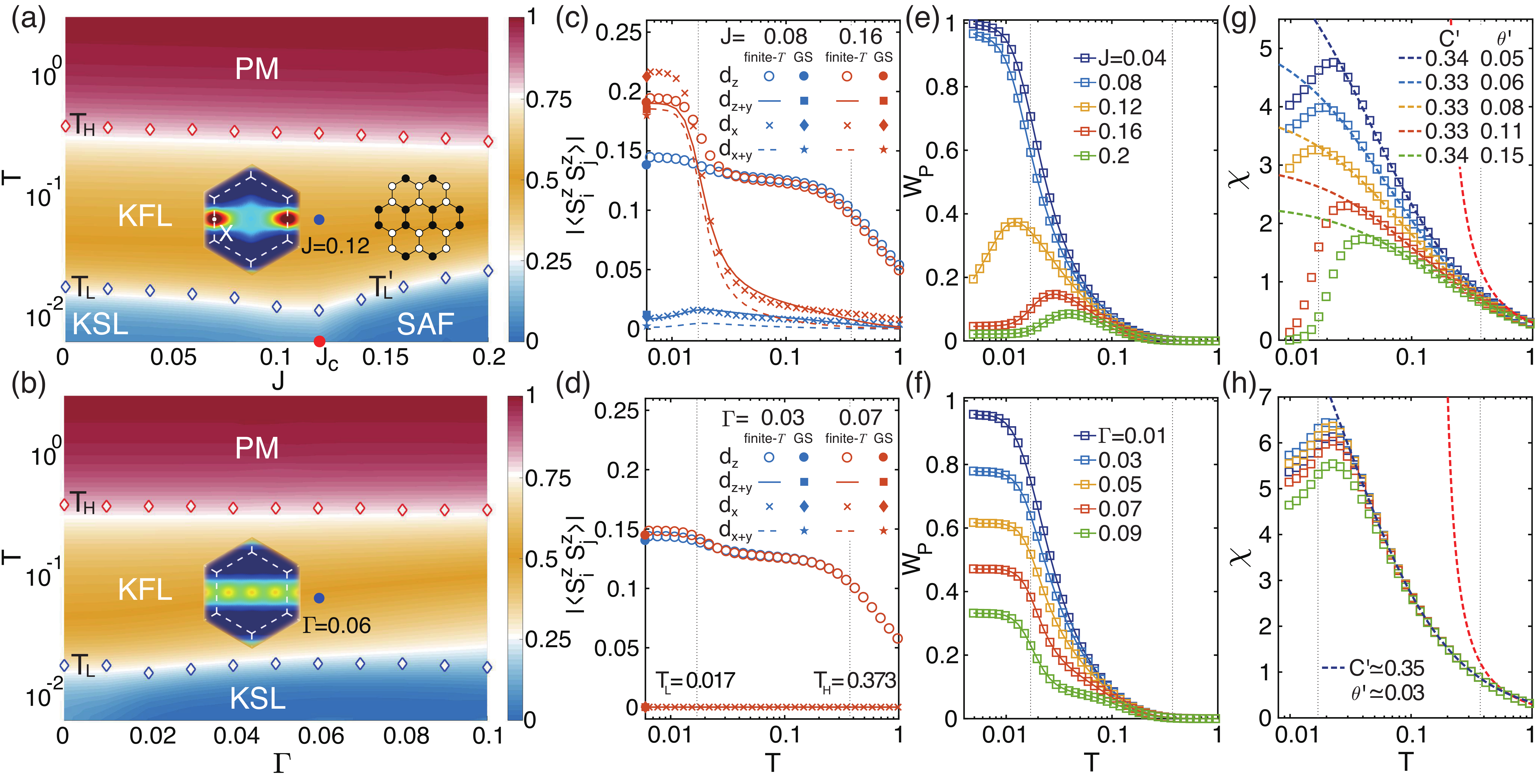}
\renewcommand{\figurename}{\textbf{Fig. }}
\caption{The phase diagram of (a) Kitaev-Heisenberg and
(b) Kitaev-$\Gamma$ model obtained by entropy $S/\ln2$,
where the white diamonds indicate the crossover temperature determined from the double-peaked specific heat.
Insets are the $T=0.07$ static structure factors of (a) $J=0.12$ and (b) $\Gamma=0.06$ in the extended Brillouin zone.
The XTRG results of the correlations $\langle S_i^z S_j^z\rangle$ along four NN bonds
with (c) $J=0.08$, $0.16$
and (d) $\Gamma=0.03$, $0.07$ all converge at lowest temperature to the
GS values calculated by DMRG.
For the Kitaev-Heisenberg with $J=0.08$ and the Kitaev-Gamma cases,
$d_z$ correlations rise up at around $T_H$ and stay predominate across
the whole temperature range; while for the $J=0.16$ case, all correlations
increases abruptly at the low temperature scale, indicating the onset of
SAF order. (e, f) Flux $W_{\rm P}$ with various $J$ or $\Gamma$, and
(g, h) magnetic susceptibility $\chi$ and the Curie-Weiss fittings for two models.
The $\chi$ data are computed under a small uniform field $h=0.002$.
}
\label{FigKHKG}
\end{figure*}

\begin{figure}[t!]
\includegraphics[angle=0,width=1\linewidth]{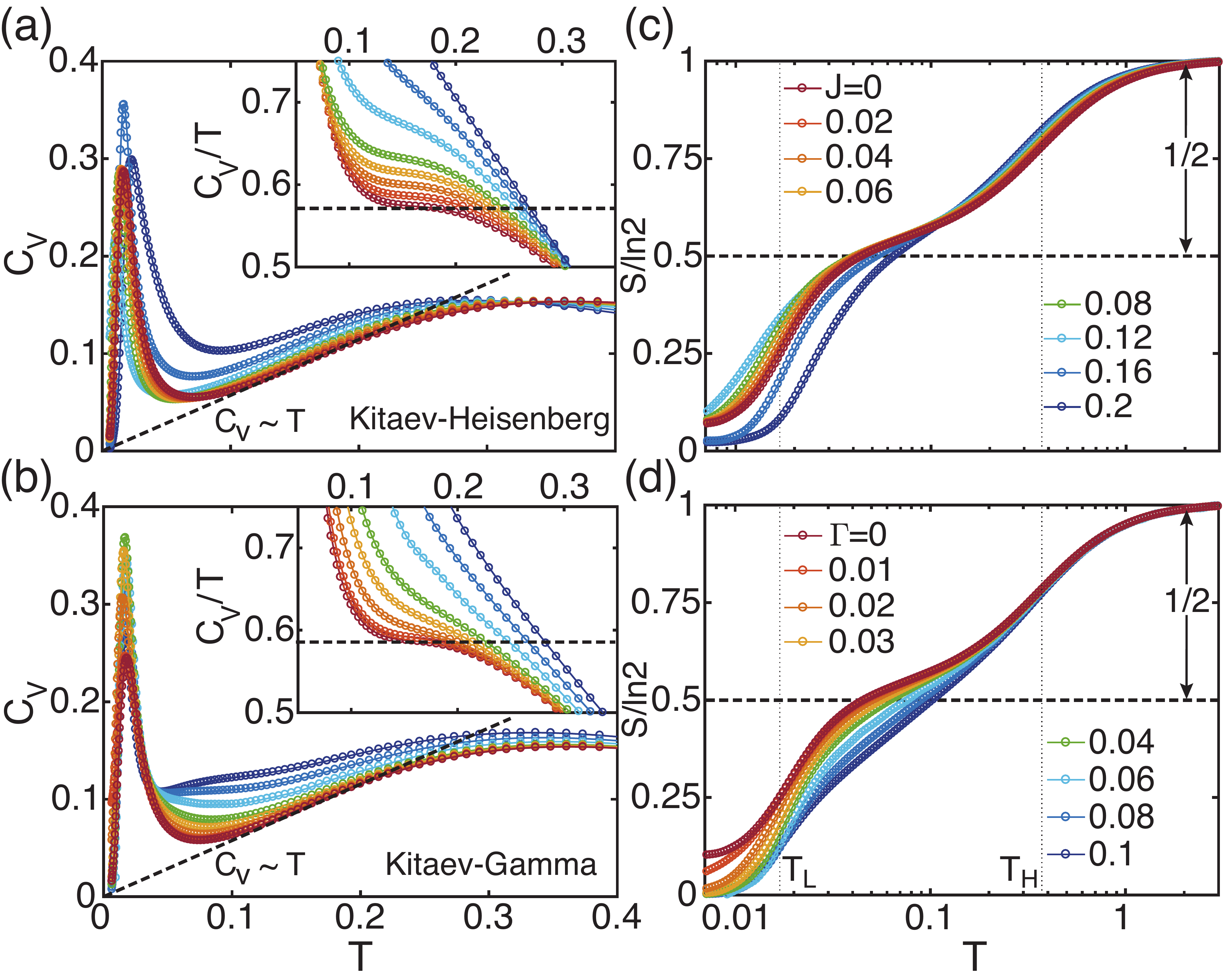}
\renewcommand{\figurename}{\textbf{Fig. }}
\caption{Specific heat $C_V$ vs. $T$ for the (a) Kitaev-Heisenberg
and (b) Kitaev-Gamma models with various perturbation strengths.
The black dashed lines indicate the linear $T$-dependence of $C_V$
at intermediate temperature, and the corresponding $C_V / T$ 
data are shown in the insets. For Heisenberg perturbation in (a),
the $C_V/T$ plateau is apparent for $J \lesssim 0.08$,
and leaves a shoulder-like structure until $J_c\simeq 0.12$;
while for the Gamma perturbations this plateau and
its remnant signature are clear for $\Gamma \lesssim 0.04$.
In (c,d) we plot the thermal entropy results of both models.
Around the stable high-$T$ scale $T_H$,
the entropy curves almost collapse for various $J$ and $\Gamma$ strengths,
and consequently leaves a quasi-plateau with $S\simeq \frac{1}{2} \ln{2}$.
}
\label{FigMetallic}
\end{figure}

We denote the energy eigenstates by $| E^{n}_{\{W_P\}} \rangle$, where $\{W_P\}$
represents the flux configurations, and $n$ labels the individual states within the $\{W_P\}$  sector.
By inserting the orthonormal basis, we arrive at the Lehmann spectral representation
\begin{widetext}
\begin{eqnarray}
\langle S_{i_0}^z(\tau) S_j^z\rangle_{\beta}  = &&  \sum_{\{W_P\},n} \sum_{n'}  e^{- \beta E^n_{\{ W_P \}}} \, e^{- \tau
\Delta_{n, \{W_P\};n',\{W'_P\}} } \, \langle E^n_{\{W_P\}} | S_{i_0}^z | E^{n'}_{\{W'_P\}} \rangle
\langle E^{n'}_{\{W'_P\}} | S_{j}^z | E^{n}_{\{W_P\}} \rangle,
\end{eqnarray}
\end{widetext}
where $| E^{n'}_{\{W'_P\}} \rangle$ represents a state $n'$
in the flux-flipped sector $\{ W'_P \}$.
Since the itinerant Majorana fermions are weakly coupled to the flux
at intermediate temperature \cite{Nasu2015}
and thus the excitation energy is close to the flux gap,
$\Delta_{n,\{W_P\};n',\{W'_P\}} =
|E^{n'}_{\{W'_P\}}-E^n_{\{ W_P \}}| \sim T_L \ll T \equiv 1/\beta$.
Therefore, the decay factor
$e^{- \tau \Delta_{n,\{W_P\};n',\{W'_P\}}} \simeq 1$, which is
not surprising as different flux sectors are nearly degenerate
in the fractional liquid regime, as evidenced by $W_P\ll 1$
[see, e.g., \Fig{FigCurves}(c)].

Consequently, $\langle S_{i_0}^z(\tau) S_j^z\rangle_{\beta}$
is virtually $\tau$-independent at intermediate temperature and can be
approximated as $\langle S_{i_0}^z S_j^z\rangle_{\beta}$ in \Eq{EqDynaSus},
thus recovering the conventional susceptibility formula
\begin{equation}
\label{EqStatSus}
\chi = \frac{1}{T} \sum_j \langle S_{i_0}^z S_j^z \rangle_\beta,
\end{equation}
which holds only for $T \gtrsim T_L$.
As observed in \Fig{FigCurves}(a) and \Fig{FigSus}(d),
the correlation $\langle S_{i_0}^z S_j^z \rangle_\beta$
exhibits a quasi-plateau vs. temperatures between $T_L$ and $T_H$,
resulting in a $1/T$-divergent susceptibility that
accounts for the emergent Curie law in the fractional liquid regime.
On the other hand, in the KSL regime at $T \lesssim T_L$,
albeit the correlation remains extremely short there,
the (flux) excitation energy $\Delta_{n,\{W_P\};n',\{W'_P\}}
\sim T_L \gtrsim T$
leads to an exponentially suppressive factor in $\tau$,
and thus a non-divergent susceptibility $\chi$ below $T_L$ in \Fig{FigSus}.

Above we have discussed the emergent susceptibility scaling in the FM Kitaev model,
and below we briefly discuss the case of AF model, and refer the details to Appendix~\ref{AppKGAFSus}.
For the AF Kitaev model, the spin correlations are also extremely short-ranged,
whose absolute values are identical to those of the FM case, but with opposite sign (AF correlation).
Therefore, the emergent Curie law is absent in the uniform susceptibility $\chi$
of the AF Kitaev model, while it appears instead in the stagger susceptibility.
This can be understood intuitively as follows:
one can flip simultaneously the spins on one of the two sublattices on the honeycomb lattice,
tuning the AF Kitaev model into an FM one, and the staggered susceptibility
$\chi_s$ corresponds to the uniform $\chi$ exhibiting universal scaling at intermediate temperature.
Such an interesting correspondence between the FM Kitaev model under uniform field
and the AF model under staggered field has been noticed in other studies under different
circumstances \cite{Trebst2019}. Our results indicate a sensitive way to determine
the sign of Kitaev coupling via analyzing the susceptibility measurements.

\section{Fractional liquid under non-Kitaev perturbations}
\label{SecKFLpert}

The Heisenberg coupling is proposed to be relevant
in the Kitaev materials Na$_2$IrO$_3$ \cite{Chaloupka2010,Singh2012},
and the Kitaev-Heisenberg model has thus been extensively studied
for both the ground-state \cite{Chaloupka2010, Chaloupka2013, Jiang2011, Schaffer2012, Gotfryd2017}
and finite-$T$ properties \cite{Reuther2011, Price2012}. 
On the other hand, in $\alpha$-RuCl$_3$ there exist a symmetric off-diagonal
exchange --- the so called Gamma term \cite{Rau2014, Gohlke2018, Juraj2019, Jiucai2019a} ---
which turns out to be important non-Kitaev interactions.
In this section, we explore the Kitaev-Heisenberg (on YC$4\times6\times2$ cylinders)
and Kitaev-Gamma (YC$4\times4\times2$)
models [See \Eq{EqHamil}] at finite temperature,
from which we see that although the low-temerapture
fate of the Kitaev model may be drastically affected
by the Heisenberg and Gamma terms,
the behavior at high and intermediate temperatures,
especially in the Kitaev fractional liquid regime,
remain qualitatively unchanged.

\subsection{Fractional thermal entropy and specific heat hehaviors}
We start with the thermal entropy and specific heat calculations
of the Kitaev-Heisenberg and Kitave-Gamma models,
and show the results in Figs.~\ref{FigKHKG} and \ref{FigMetallic}.
Between the two entropy-release steps [\Fig{FigMetallic}(c,d)],
there exists an extended regime with thermal
entropy $S \simeq \frac{1}{2}\ln2$ [\Fig{FigKHKG}(a,b)]
for both models, which indicate the existence of
Kitaev fractional liquid at intermediate temperature.
The 1/2-entropy quasi-plateau remains to be evident
till Heisenberg coupling $J\lesssim0.16$ and for
Gamma perturbation of $\Gamma \lesssim 0.06$.
Beyond this $\Gamma$ or $J$ range,
a shoulder-like feature with slow entropy release
can still be recognized in both models.
In \Fig{FigKHKG}(a,b), the crossover temperature scales, $T_H$ and $T_L$ ($T_L'$),
can be determined from the isentropic lines of $S/\ln2=0.75$ and $0.25$, which
coincide with the higher and lower crossover temperatures
in the double-peaked specific heat curves
[see \Fig{FigMetallic}(a,b) and also Appendix~\ref{AppThermoKHKG}].

In \Fig{FigMetallic}(a,b), we show that the linear-$T$ behavior
of specific heat also survives  in the presence of a finite $J$ or $\Gamma$ perturbation,
which can be also recognized from the $C_V/T$ plots in the insets.
This universal specific heat behavior suggests
the emergence of Majorana-metal states
in the fractional liquid regime.
Nevertheless, similar to what was observed in \Fig{FigCv},
the $J$ and $\Gamma$ perturbations can shift the
flux excitations as well as other low-energy spin fluctuation modes
towards higher temperatures and thus may mask the linear-$T$ $C_V$,
leaving eventually only a shoulder-like reminiscent
in the insets of \Fig{FigMetallic}(a,b).

The finite-$T$ properties can sensitively reflect the quantum
phase transition in the ground state. In \Fig{FigKHKG}(a),
$T_L$($T'_L$) is non-monotonic vs. $J$:
For $J \lesssim J_c \simeq 0.12$, $T_L$ represents
a crossover from the KSL to the fractional liquid; 
while for $J>J_c$, the stripe anti-ferromagnetic (SAF) order sets in below $T'_L$,
where correlations along $d_x$, $d_{x+y}$, etc,
also increase rapidly as shown in \Fig{FigKHKG}(c).
Therefore, we infer from the finite-$T$ results that
the Kitaev-Heisenberg model undergoes a
transition from the KSL to SAF phase at around $J_c \simeq 0.12$,
consistent with previous ground-state studies
\cite{Chaloupka2010, Schaffer2012, Jiucai2019a}.

\subsection{Bond-directional spin correlation and stripy structure factor}
\label{sec:spincorr}
From the spin-correlation data shown in \Fig{FigKHKG}(c,d),
we emphasize that the two models share common features
in the Kitaev fractional liquid regime. First, the spin correlation remains extremely short-ranged,
and mostly present on the $d_z$ bonds. This is consistent with a perturbation-theory analysis:
except on the $d_z$ bonds, spin correlations are suppressed because $\mathbb Z_2$
fluxes are approximately conserved. Furthermore,
such correlations appear
at the first order in $J$ but only at the second order in $\Gamma$,
consistent with the fact that $J$ has a larger effect on
spin correlations than $\Gamma$, as shown in Fig.~\ref{FigKHKG}(c,d).
This prominent spin-correlation feature consequently leads
to the stripy structure factor in the insets of Fig.~\ref{FigKHKG}(a,b).

Under the Heisenberg perturbations, we observe
in the inset of \Fig{FigKHKG}(a) a clear structure peak at the X point,
which indicates the enhanced short-range SAF correlation,
with the classical spin pattern shown rightside.
Besides the SAF peak, we notice that there exists a stripy
structure background which is related to
the predominate bond-directional NN correlations at intermediate temperature.
On the other hand in \Fig{FigKHKG}(b), for the Kitaev-Gamma model,
the low temperature scale $T_L$ only slightly changes.
The KSL-like stripy spin structure can be observed with no clear signature of spin ordering,
which, together with the nonzero fluxes, $W_P \neq 0$, at low temperature in \Fig{FigKHKG}(f),
suggest that the low-$T$ quantum state remains in the asymptotic KSL phase
at least for $\Gamma \lesssim 0.1$, in agreement with recent ground-state
studies of the Kitaev-Gamma model \cite{Gohlke2018,Jiucai2019a}.

\begin{figure}[t!]
\includegraphics[angle=0,width=0.95\linewidth]{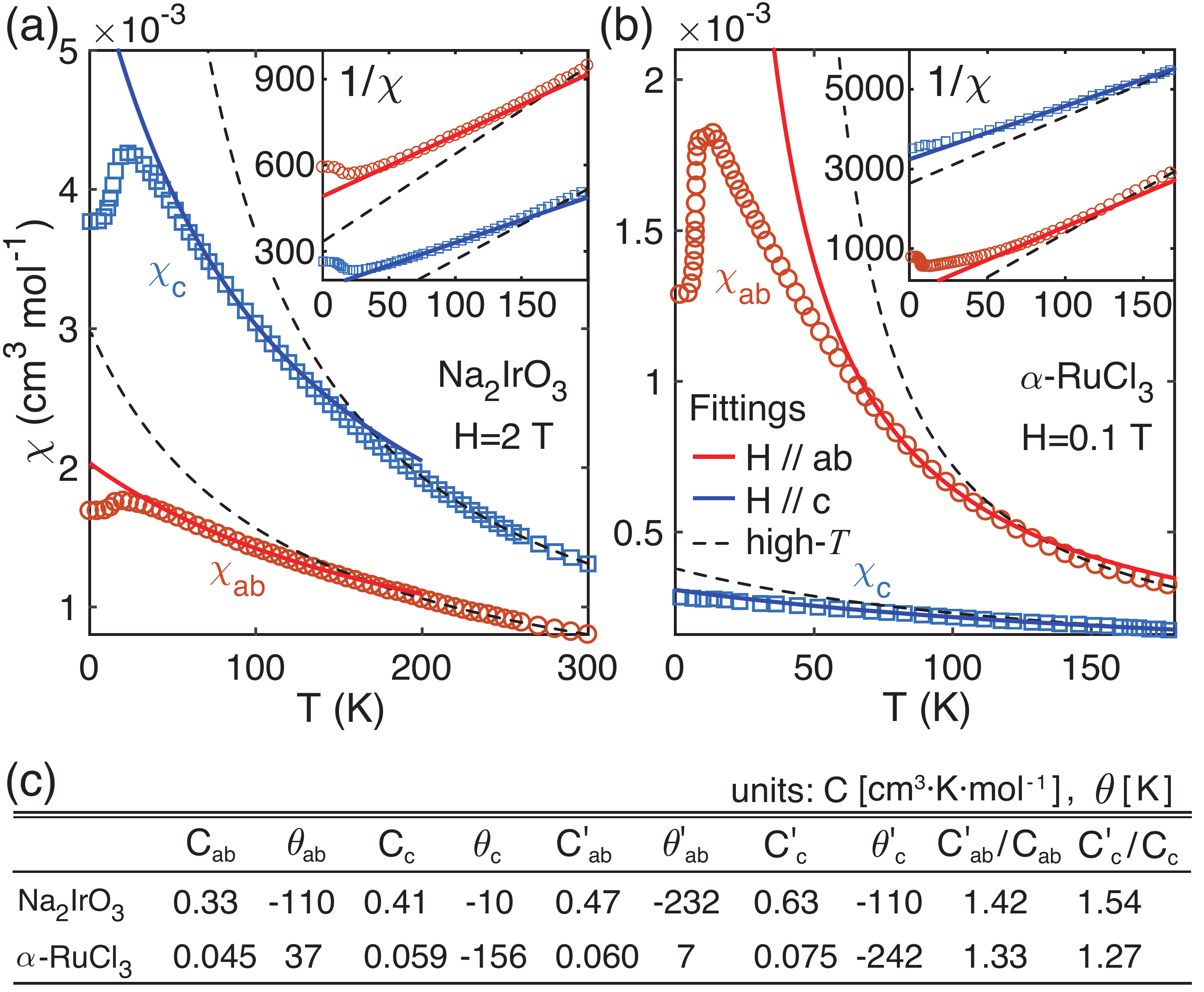}
\renewcommand{\figurename}{\textbf{Fig. }}
\caption{(a, b) Experimental Na$_2$IrO$_3$ \cite{Singh2010} and $\alpha$-RuCl$_3$
susceptibility data \cite{Kutoba2015}
and their Curie-Weiss fittings at both high and intermediate temperatures
under fields H // $ab$ (orange circles) and H // $c$ (blue squares), with
inverse susceptibility plotted in the insets. (c) The fitted Curie constants ($C$ for high- 
and $C'$ for intermediate-$T$), and the Curie temperatures 
(corresponding $\theta$ and $\theta'$), 
for both $c$ and $ab$ directions, are listed in the table. 
}
\label{FigSusKMater}
\end{figure}

\subsection{Universal susceptibility scaling}
Now we switch to the analysis of the magnetic susceptibility scalings in both models.
At intermediate temperature, we again observe bond-directional spin correlations
that are mostly restricted within the NN sites [see \Fig{FigKHKG}(c,d) and discussions in Sec.~\ref{sec:spincorr}],
as well as the strongly fluctuating flux [\Fig{FigKHKG}(e,f)]
under an extended range of $J$ or $\Gamma$ perturbations.
Following the line of discussions in the pure Kitaev model, we expect an
emergent Curie law of susceptibility in the fractional liquid,
and this is indeed observed in \Fig{FigKHKG}(g,h).
Although the temperature window of the emergent Curie behaviors
gets narrowed as $J$ or $\Gamma$ increases, the universal
susceptibility scaling can be very clearly identified across the
intermediate-$T$ regime, regardless of whether the ground state is a KSL
or semiclassically ordered.

Nevertheless, there are still some difference in the susceptibility scalings between the Heisenberg and
Gamma perturbations in the susceptibility scalings. At the temperature above $T_L$,
the susceptibility curves in the Kitaev-Gamma model collapse into the same universal behavior,
with $C' \approx 1/3$ (thus $C'/C \approx 1.33$, the same to the pure Kitaev model)
and with the same small $\theta'$. This is in contrast to the case of the Kitaev-Heisenberg model,
where although $C'\simeq1/3$ remains the value of the pure Kitaev model,
the Curie-Weiss temperature $\theta'$ grows as $J$ increases, which suggests
the fluxes are no longer ``free'' but coupled with each other
since the Heisenberg perturbation $J$ introduces longer-range correlations
to the system [\Fig{FigKHKG}(c)] .
Accordingly in \Fig{FigKHKG}(e), the gauge flux for $J\lesssim J_c$ behaves
very similarly to that of the pure Kitaev model,
i.e., $W_{\rm P}$ freeze at around $T_L$;
while it gets suppressed at low temperature for $J>J_c$.

Overall, we observe that the emergent Curie law of susceptibility scaling appears
to be robust against the perturbations at intermediate temperature,
which is found to be suppressed only when the finite flux gap or the longer-range correlation
induced by the non-Kitaev terms takes effects at the low temperature scale.
Therefore, such universal susceptibility scaling can serve as a useful tool
probing the fractional liquid in the Kitaev materials.

\section{Diagnosis of the Kitaev Fractional Liquid}
\label{SecSusKMater}
\subsection{Kitaev paramagnetism in the Kitaev materials}
Thermodynamic features, including the double-peaked specific heat 
curve, linear-$T$ behavior, and the fractional thermal
entropy $S$ integrated from $C_V/T$ have been exploited in
detecting finite-$T$ fractionalization in the Kitaev materials
\cite{Kutoba2015,Yamaji2016,Do2017,Mehlawat2017,Widmann2019}. 
However, 
the magnetic response to external fields --- the susceptibility $\chi$ --- has been much less analyzed,
except for the high-$T$ behaviors in the Kitaev materials.
With insight from the emergent intermediate-$T$ Curie susceptibility
from the above model studies, we revisit the experimental $\chi$ data of two prominent Kitaev materials,
i.e., Na$_2$IrO$_3$ \cite{Singh2010} and $\alpha$-RuCl$_3$ \cite{Kutoba2015},
which are believed to realize the Kitaev model under the
Heisenberg and/or Gamma perturbations.

\begin{figure*}[t!]
\includegraphics[angle=0,width=1\linewidth]{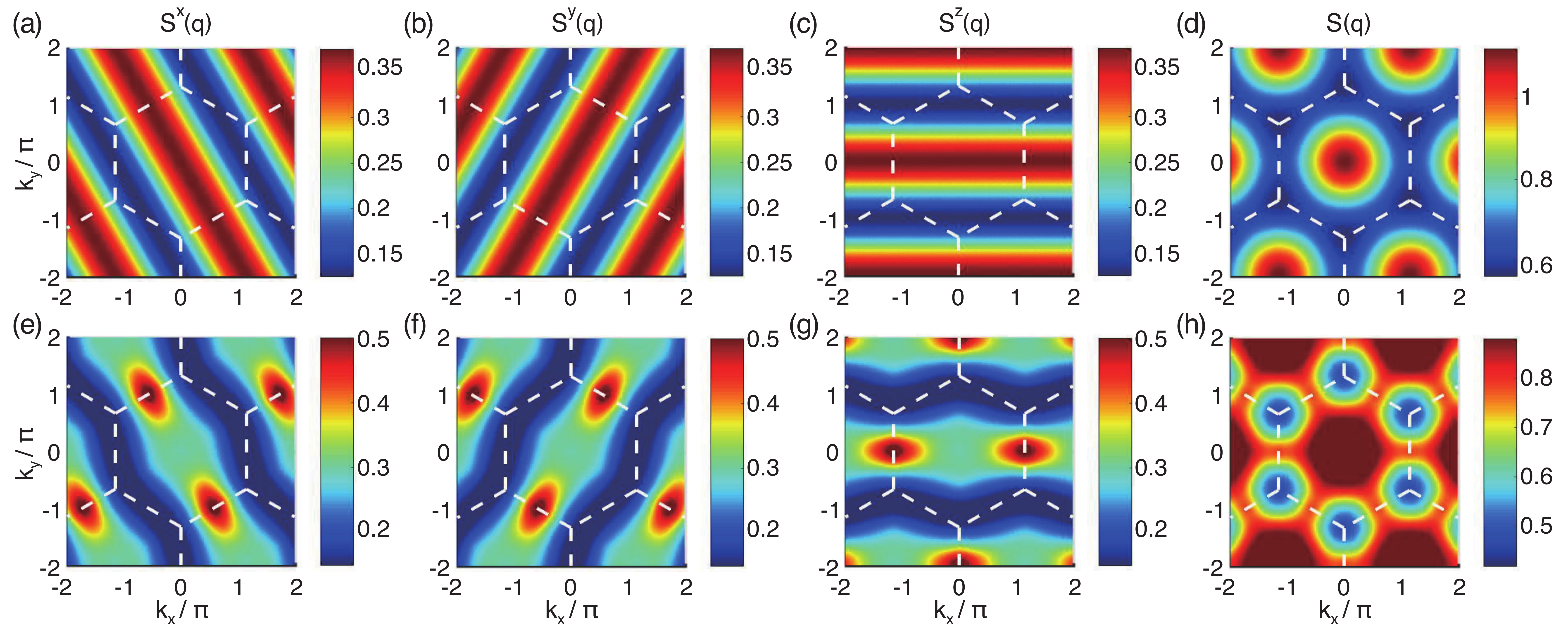}
\renewcommand{\figurename}{\textbf{Fig. }}
\caption{The static spin structure factors $S^\gamma(q)$
calculated from the equal-time spin correlation functions [see \Eq{EqSq}] 
at an intermediate temperature $T\simeq0.15$.
(a-d) exhibit the pure Kitaev model results of three spin structure factors 
$S^{x,y,z}(q)$ as well as their sum $S(q)$, and (e-h) present results of
the Kitaev-Heisenberg model with $J=0.14$.
}
\label{FigSqMap}
\end{figure*}

In \Fig{FigSusKMater},  we analyze the in-plane
and out-of-plane susceptibility
data of both materials, where the emergent Curie
behaviors can be observed right below certain
high temperature scales, i.e.,
$T_H^{\rm Ir} \simeq 120$~K and $T_H^{\rm Ru} \simeq 110$~K
for Na$_2$IrO$_3$ and $\alpha$-RuCl$_3$, respectively.
$T_H^{\rm Ir}$ and $T_H^{\rm Ru}$ are in agreement with
the higher crossover temperatures in the magnetic specific heat curves of two materials.
For example, in Na$_2$IrO$_3$ the higher specific heat
peak locates at around 110-120~K \cite{Mehlawat2017, Kim2015}
which is in very good agreement
with the $T_H^{\rm Ir}$ in \Fig{FigSusKMater}(a).
Similarly, the high-$T$ entropy rapidly releases
at around 100-110~K in
$\alpha$-RuCl$_3$ \cite{Kutoba2015,Do2017,Widmann2019},
also coinciding with $T_H^{\rm Ru}$ in \Fig{FigSusKMater}(b).

Through Curie-Weiss fittings of the intermediate-$T$
susceptibility results in \Fig{FigSusKMater},
we find the the ratios between intermediate- and high-$T$
Curie constants $C'/C \sim 1.3$ in both materials
(and either within the $ab$-plane or parallel to the $c$ axis),
suggesting the existence of the Kitaev fractional liquid
in these two Kitaev materials.
The emergent diverging susceptibility at intermediate temperature
is very sensitive to the sign of Kitaev interaction $K$,
and for the AF case ($K>0$) there exhibits a convex uniform susceptibility
curve without any emergent Curie behavior at intermediate temperature (Appendix~\ref{AppKGAFSus}).
Therefore, the analysis in \Fig{FigSusKMater} suggests
the FM Kitaev interactions in both compounds
Na$_2$IrO$_3$ and $\alpha$-RuCl$_3$,
in agreement with previous theoretical \cite{Yamaji2014,Winter2016,Janssen2017,Winter2017}
and experimental \cite{Sears2020} estimates.

Meanwhile, in \Fig{FigSusKMater} we observe the
intermediate Curie-Weiss behaviors are deviated
at the lower temperature $T_L^{\rm Ir} \simeq 50$-55~K and
$T_L^{\rm Ru} \simeq 60$-65~K. Therefore,
in these two materials there are fractional liquid regimes
stretching appreciable temperature ranges as wide as 65-70~K (Na$_2$IrO$_3$)
and 45-50~K ($\alpha$-RuCl$_3$), respectively.
In addition, by comparing \Fig{FigSusKMater}(a) to \Fig{FigKHKG}(g),
we find a relatively wider intermediate-$T$ range
for the emergent Curie-Weiss in Na$_2$IrO$_3$.
This resembles that of the Kitaev-Heisenberg model,
where universal scaling is deviated at around the susceptibility peak temperature.
This observation suggests that indeed the Heisenberg-type perturbation
can play an important role in this material \cite{Chaloupka2010}.
On the other hand, in \Fig{FigKHKG}(g) the Gamma term deflect the
Curie-Weiss behavior way before the susceptibility peaks,
and this bears a resemblance to the
case of $\alpha$-RuCl$_3$ shown in \Fig{FigSusKMater}(b),
supporting thus the proposal that the $\Gamma$ term (together with other perturbations)
is particularly relevant in this material \cite{Ran2017,Liu2018}.

Overall, through analysis of the magnetic susceptibility data of Na$_2$IrO$_3$
and $\alpha$-RuCl$_3$, we find a universal susceptibility scaling that may be ubiquitous
in other members of the intriguing family of Kitaev materials and can serve
as a sensitive thermodynamic signature of the Kitaev fractional liquid.
Besides, the susceptibility scaling indicates the existence of
ferromagnetic Kitaev interactions, together with other non-Kitaev couplings,
in the two prominent Kitaev materials.

\subsection{Spin-correlation diagnostics of the Kitaev fractional liquid}
\label{Subsec:SCD}
The short-range correlated Kitaev fractional liquid leaves unique features
in the spin correlations and thus structure factors. At intermediate temperature,
the nonzero correlation $\langle S_i^\gamma S_j^\gamma \rangle$ is
restricted within the NN $\gamma$-bond,
with the rest correlations negligibly small, even under magnetic fields
and non-Kitaev perturbations, as shown in Figs.~\ref{FigCurves}(a,b) and \ref{FigKHKG}(c,d).
Correspondingly, in Figs.~\ref{FigPhsDigFld}(c) and \ref{FigKHKG}(a,b)
we see this peculiar bond directional correlation pattern results
in a stripy-like background in the $z$-direction spin structure factors $S^z(q)$.

Matter of fact, this stripy spin structure also appears
in the structure factors $S^\gamma(q)$ with $\gamma=x,y$, and the direction
is ``locked'' with the associated spin components $\gamma$.
As shown in \Fig{FigSqMap}(a,e), the $S^x(q)$ stripe extends from the left-top
to the right-bottom, which rotates counterclockwise by $120^{\circ}$
when switched to $S^y(q)$ [\Fig{FigSqMap}(b,f)],
followed by another $120^{\circ}$ rotation to $S^z(q)$ in \Fig{FigSqMap}(c,g).
Note that the spin structure factors results in \Fig{FigSqMap} are evaluated
at an intermediate temperature $T\simeq0.15$
for the pure Kitaev model (the upper row) and Kitaev-Heisenberg model (lower).
In the latter Kitaev-Heisenberg case with $J=0.14$, there emerges
a structure factor peak representing the semicalssical SAF correlation,
which also rotate as the spin component switches. Such a rotation of short-range correlation peak
has previously been observed in the spin-resolved magnetic X-ray measurements
right above the ordering transition temperature ($T\simeq 17$~K) \cite{Kim2015},
which provides a direct evidence of the dominant bond-directional interactions
in the Kitaev material Na$_2$IrO$_3$.
Therefore, our findings in \Fig{FigSqMap}(e-g)
confirm that this experimental observation is indeed related
to the presence of Kitaev interactions in the material,
although the SAF order involved here is different from the
zigzag type in Na$_2$IrO$_3$.

Moreover, our results in \Fig{FigSqMap}(d,h) also
shed light on the renowned star-like spin structure factor
observed in inelastic neutron scattering measurement of $\alpha$-RuCl$_3$ \cite{Banerjee2017,Do2017}.
From the full structure factor $S(q) =  \sum_\gamma S^\gamma(q)$,
we observe in the pure Kitaev model that the superposition of the three stripes
leads to an approximately rotationally symmetric shape in \Fig{FigSqMap}(d).
By adding bright spots representing short-range SAF correlations on top of that,
we instead see a six-fold symmetric $S(q)$ in \Fig{FigSqMap}(h).
This clearly suggests that the star-shape structure factor (with six-fold symmetry) observed
in inelastic neutron scatterings on $\alpha$-RuCl$_3$ can be ascribed to the
coexistence of the Kitaev bond-directional correlation and the short-range zigzag order.
As the temperature/energy scale gradually increases, it approaches asymptotically the rotationally
symmetric spin structure in \Fig{FigSqMap}(d), which has also been observed in experiments.

In general, we propose that the bond-directional spin correlation and related
intermediate-$T$ stripy feature provide a useful tool for diagnosing the Kitaev materials.
Specifically, diffuse X-ray (for Na$_2$IrO$_3$)
or spin-resolved neutron (for $\alpha$-RuCl$_3$) measurement
at intermediate temperature, say, $T=50$-100~K, can be exploited
to observe the stripy spin structures $S^\gamma(q)$ that serves
as a direct evidence of the Kitaev fractional liquid.

\section{Discussions and Outlook}
\label{SecConOut}
In conventional spin models and magnetic materials,
a ``crystallization'' of spins occurs spontaneously, 
from the high-$T$ gas-like paramagnetic to the low-$T$ phase
with long-range order. Interestingly, in the frustrated 
Kitaev model and related Kitaev materials there exists an alternative scenario, 
where a fractionalization of spins takes place upon cooling. 
This gives rise to an intermediate-$T$ Kitaev fractional liquid
and eventually to the low-$T$ quantum spin liquid phase
with long-range entanglement.

Exploiting the XTRG method in the finite-$T$ simulations,
aided by the ground-state DMRG calculations,
we reveal unambiguously a Kitaev fractional liquid 
in the Kitaev model under perturbations, including the magnetic field,
Heisenberg and Gamma terms. In the exotic finite-$T$ quantum states, 
there exist fractional thermal entropy, linear-$T$ specific heat,  
and, in particular, an emergent Curie law in magnetic susceptibility.
These thermodynamic features are robust against perturbations, 
regardless if the ground state is still in a spin liquid phase or not.

Remarkably, this universal Curie susceptibility that reflects the thermally
free flux degrees of freedom, together with the previously
recognized metallic specific heat behaviors of itinerant Majorana,
each revealing one aspect of the spin fractionalization,
constitute a comprehensive thermodynamics characterization of
the exotic Kitaev fractional liquid. In light of this, and from analyzing
the experimental susceptibility data, we find signatures of the Kitaev 
fractionalization liquid in two Kitaev materials
Na$_2$IrO$_3$ and $\alpha$-RuCl$_3$, which support 
ferromagnetic Kitaev interactions in the two materials.  

Matter of factor, the diverging susceptibility is a robust thermodynamic signature
also related to the extremely short-ranged and bond-directional spin correlations 
in the fractional liquid regime, which corresponds to a stripy spin structure factor.
Therefore, besides with the magnetic susceptibility measurement,
our findings also encourage further material exploration
of the Kitaev fractional liquid using spin-resolving neutron and resonant X-ray scatterings.
Lastly, our work paves the way for the precise determination of interaction parameters 
in the Kitaev materials by fitting the measured thermodynamic data, which we leave in a future study.

\begin{acknowledgments}
\textit{Acknowledgments.---}
The authors are indebted to Zheng-Xin Liu, Jiu-Cai Wang, Hao-Xin Wang, 
Andreas Weichselbaum, Wentao Jin, and Jan von Delft for insightful discussions.
This work was supported by the National Natural Science Foundation of China
(Nos. 11974036, 11874078, 11834014, and 11874115),
and the Fundamental Research Funds for the Central Universities.
\end{acknowledgments}

\appendix

\setcounter{figure}{0}
\renewcommand\thefigure{A\arabic{figure}}

\section{Exponential Tensor Renormalization Group Method Applied to the perturbed Kitaev Model}
\label{App:XTRG}

\begin{figure}[t!]
\includegraphics[angle=0,width=1\linewidth]{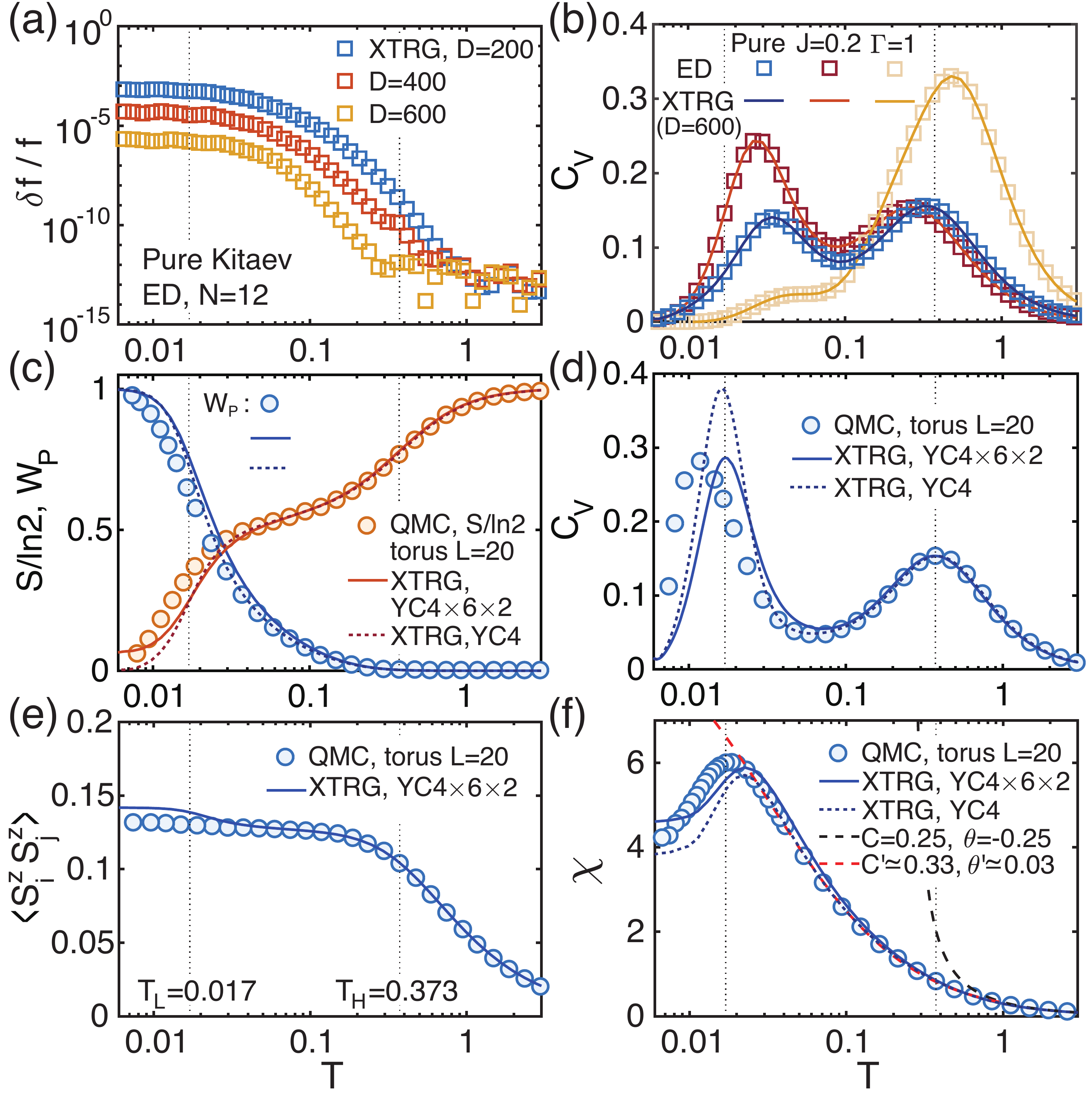}
\renewcommand{\figurename}{\textbf{Fig.}}
\caption{(a) The relative errors $\delta f / f$ between the ED and XTRG
results of a isotropic pure Kitaev model on the $N=12$ (YC4) cluster,
with retained bond dimensions $D=200$, $400$ and $600$.
(b) The corresponding specific heat results of pure Kitaev (blue),
Kitaev-Heisenberg with $J=0.2$ (red),
and Kitaev-Gamma model with $\Gamma=1$ (yellow).
Comparisons between the large-scale XTRG and QMC simulations \cite{Motome2020}
are presented for (c) the thermal entropy $S/\ln2$ (orange) and flux $W_{\rm P}$ (blue),
(d) specific heat $C_V$, and (e) spin correlations $\langle S_i^z S_j^z \rangle$.
We also compare the XTRG susceptibility results to the QMC data on the
$L=20$ torus \cite{Yoshitake2017b}, where the high- and intermediate-$T$
fittings are also performed to the latter (dashed lines).
We show in (c-f) both the YC$4\times6\times2$ results (solid curves)
and the YC4 data (dotted curves) obtained by subtracting length $L=6$ by $L=4$ results.
XTRG results of susceptibility are computed under [0 0 1] field,
which agree excellently with the QMC results in (f) above the low temperature $T_L$,
although the latter is along the [1 1 1] direction.
}
\label{FigBnchmk}
\end{figure}

The exponential tensor renormalization group (XTRG) method
reaches low-$T$ thermal states exponentially fast,
which has been shown to be highly efficient in both critical
quantum chains and various 2D lattice systems \cite{Chen2018,Chen2018b,Lih2019},
and is exploited to study the Kitaev model of interest.
To start the XTRG procedure, we needs to invoke a quasi-1D mapping
of the lattice model, and encode the ``long-range'' interaction information
in a compact matrix product operator (MPO) form.
For the pure Kitaev model on the YC4 geometry, the MPO has a bond
dimension of $D_H = 7$, which increases to $D_H=14$ for the
Kitaev-Heisenberg and Kitaev-Gamma Hamiltonians.

Given the compact MPO representation of the Hamiltonian $H$,
we can represent $\rho(\tau)$, initiated at a very small initial $\tau \sim 10^{-3 \sim -4}$,
also as an MPO conveniently and accurately (up to machine precision)
via the series expansion
\begin{equation}
\rho(\tau) = e^{- \tau \hat H} = \sum_{n=0}^{N_c} \frac{(- \tau)^n}{n!} H^n.
\end{equation}
Through this expansion, $H^n$ can also be expressed as an MPO
using tensor compression techniques \cite{Chen.b+:2017:SETTN},
followed by a sum-and-compression procedure
that finally leads to the compact MPO form of $\rho(\tau)$.

Following the XTRG idea of exponential cooling procedure,
we square the density matrix and obtain at the $n$-th step
\begin{equation}
\rho(2^{n+1} \tau) =  \rho(2^n \tau) \cdot \rho(2^n \tau),
\end{equation}
with which we can compute the free energy at $\beta \equiv 2^{n+2} \tau$
via the thermo-field double formalism
\begin{equation}
f(\beta) = -\frac{1}{\beta} \ln{ \rm{Tr} [\rho(2^{n+1} \tau) \cdot \rho(2^{n+1} \tau)]},
\end{equation}
by a bilayer tensor trace \cite{Dong.y+:2017:BiLTRG}.
Given the free energy, other thermodynamic quantities including the
specific heat, thermal entropy, and magnetic susceptibility, etc,
can also be computed.

We first benchmark the XTRG results with
exact diagonalization (ED) on a $12$-site cluster
(with the same width and boundary condition as YC4
geometry but has only three sites along the zigzag edge).
In \Fig{FigBnchmk}(a), the relative errors of free energy
$\delta f/f$ reduces as the retained bond dimension $D$ increases,
reaching a very high accuracy with $D=600$.
On top of that, we show the XTRG results of specific heat
computed with $D=600$, of pure Kitaev, Kitaev-Heisenberg,
and Kitaev-Gamma models, and show them in \Fig{FigBnchmk}(b),
where we also find excellent agreement with ED results.

As shown in \Fig{FigBnchmk}(c-f), we compute
the thermal entropy, flux expectation, specific heat,
spin correlations and magnetic susceptibility, and compare our XTRG results
with the QMC data taken from Refs.~\onlinecite{Motome2020, Yoshitake2017b}.
Despite the very different system sizes, i.e., YC4 cylinder vs. $L=20$ torus,
excellent agreement between two methods can be observed
at high and intermediate temperatures, covering the fractional liquid regime.
The modest deviation of XTRG from QMC data near the lower
temperature scale $T_L$ can be ascribed to the cylindrical boundary condition
and the finite width, which nevertheless does not lead to
any qualitative difference for our discussion.

Moreover, we have also analyzed the large-scale QMC susceptibility data in \Fig{FigBnchmk}(f),
and from the fitting we find again an emergent Curie behavior clearly at intermediate temperature.
The Curie constant $C'\simeq1/3$ is also very close to
the YC4 XTRG results in \Fig{FigSus}(a) of the main text.

\begin{figure}[t!]
\includegraphics[angle=0,width=1\linewidth]{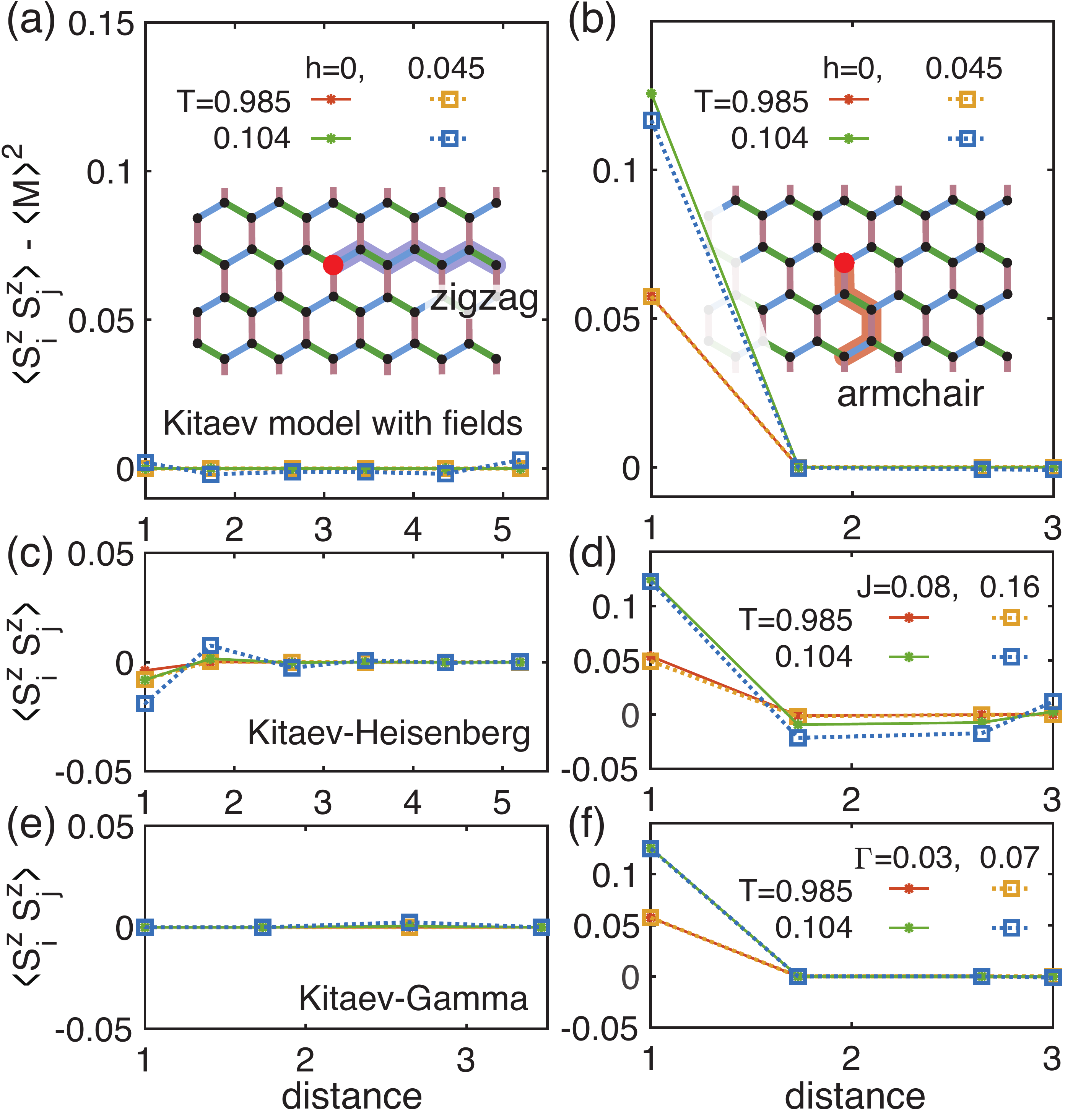}
\renewcommand{\figurename}{\textbf{Fig.}}
\caption{High- and intermediate-$T$ spin correlation results,
of various perturbed Kitaev model,
along the (a, c, e) zigzag and (b, d, f) armchair paths,
with the reference site marked in red and the hexagon edge length
as the unit of distance.
Note that in panels (a,b), we have subtracted the (field-induced)
mean magnetization squared $\langle M \rangle^2$
from the spin correlations.
}
\label{FigCorrFunc}
\end{figure}

\begin{figure}[t!]
\includegraphics[angle=0,width=0.78\linewidth]{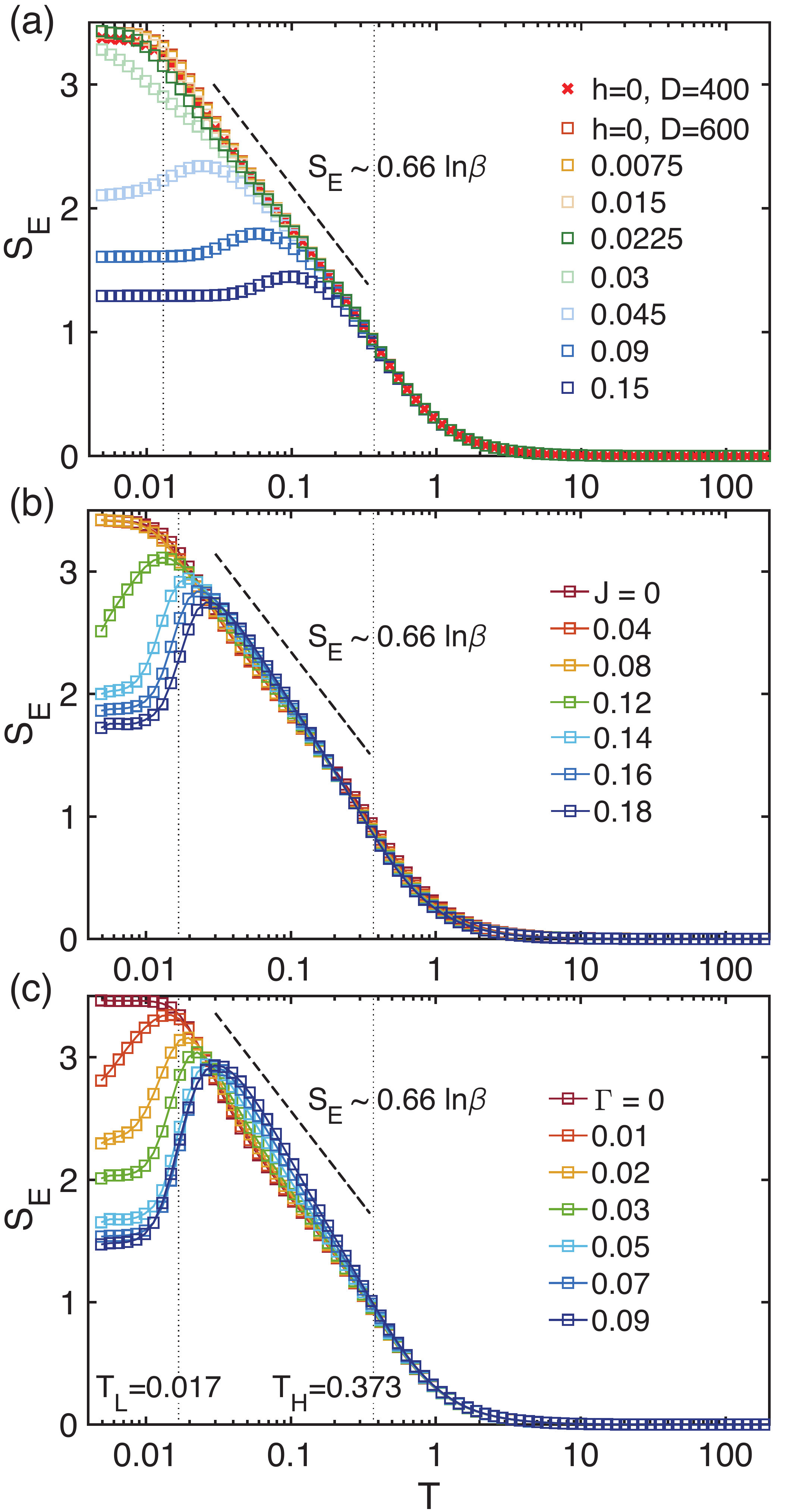}
\renewcommand{\figurename}{\textbf{Fig.}}
\caption{Bipartite entanglement $S_E$ cut at
the center of the MPO representation of density matrices
for (a) the Kitaev model with fields, (b) Kitaev-Heisenberg model, and
(c) the Kitaev-Gamma model. At intermediate temperature, $S_E$ roughly scales as
$\sim 2/3 \ln{\beta}$ which characterize the required computational resource
of the simulations.
}
\label{FigEnt}
\end{figure}

\begin{figure}[t!]
\includegraphics[angle=0,width=1\linewidth]{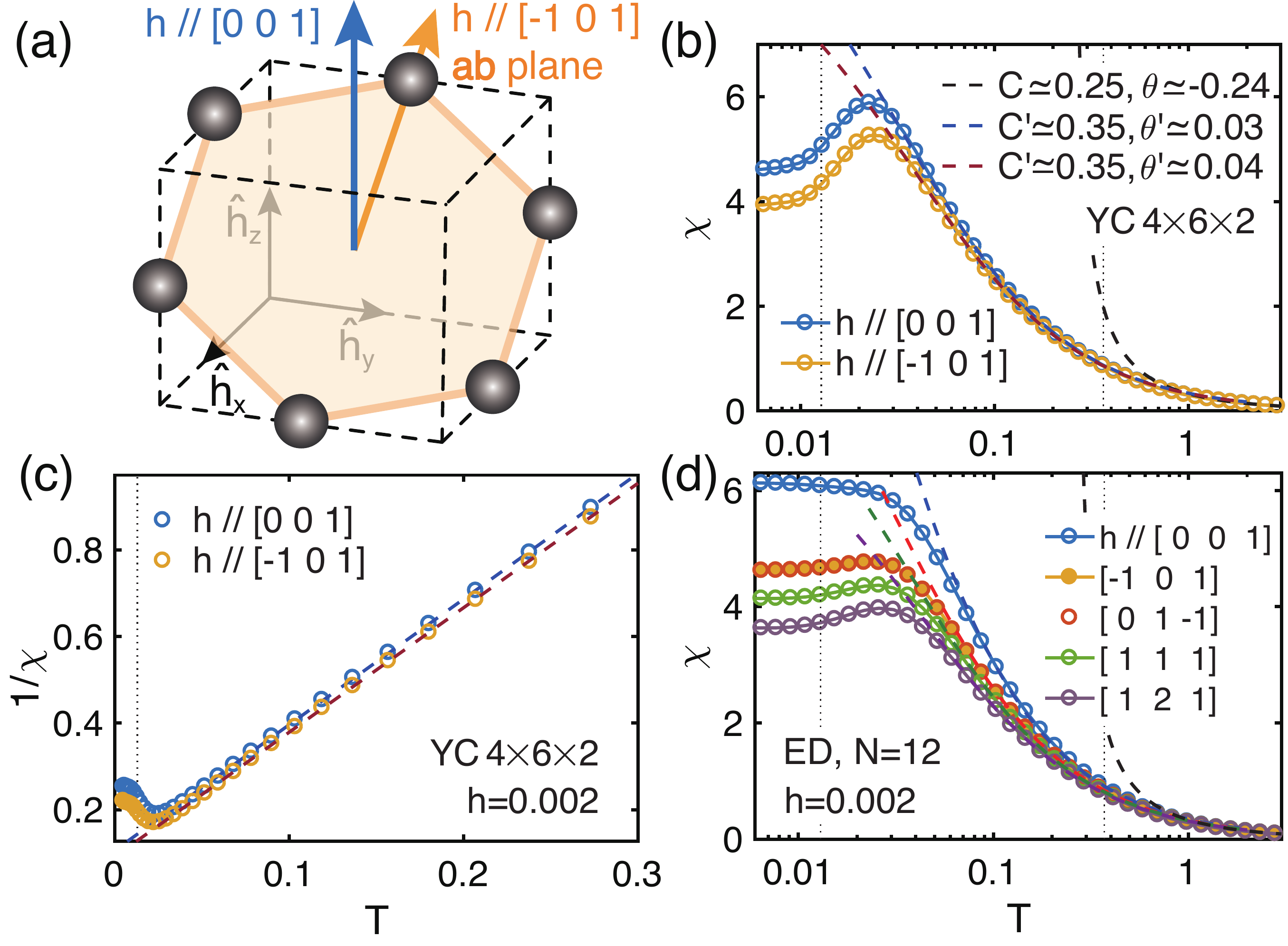}
\renewcommand{\figurename}{\textbf{Fig.}}
\caption{(a) Illustration of the magnetic fields $h$ along [0 0 1]
and [-1 0 1] (parallel to the honeycomb \textit{ab}-plane).
(b, c) The magnetic susceptibility
$\chi$ and its inverse $1/\chi$, 
calculated
with a small field $h=0.002$ .
(d) The susceptibility results along various field directions,
obtained by ED calculations on the $N=12$ cluster.
}
\label{FigVarFields}
\end{figure}

\section{Spin Correlations in the Fractional Liquid}
\label{App:CorrFunc}
The real-space spin correlations of the Kitaev model under
various perturbations are computed at a high ($T\simeq 1$) and an intermediate temperature ($T\simeq 0.1$).
In  \Fig{FigCorrFunc}(a,b), we provide $\langle S_i^z S_j^z\rangle$ along
two different (zigzag and armchair) paths, under the influence of external fields,
from which we find the correlations are bond directional and extremely short-ranged.
To be specific,  $\langle S_i^z S_j^z\rangle$ vanishes along the zigzag path in \Fig{FigCorrFunc}(a),
since there are only $x$- and $y$-bonds there. On the other hand, in \Fig{FigCorrFunc}(b),
the armchair correlations decay instantaneously for distances exceeding one,
and the nearest $z-bond$ correlation is enhanced as temperature decreases
from $T\simeq1$ to $T\simeq0.1$. For the Kitaev-Heisenberg
and Kitaev-Gamma models in \Fig{FigCorrFunc}(c-f), we observe very similar
real-space correlation features as those in \Fig{FigCorrFunc}(a,b), except
for the curves with $J=0.16$ and $T\simeq0.1$ in \Fig{FigCorrFunc}(d).
There the system enters SAF-ordered phase, where the zigzag correlations
develops a small values and the signs alternates following the SAF pattern
as shown in the inset of \Fig{FigKHKG}(a).

From the results in \Fig{FigCorrFunc}, we find
spatially anisotropic and extremely short-ranged spin correlations at intermediate temperature.
This, together with the distinct stripy spin structure factors in the main text,
provide fingerprint signatures of fraction liquid.

\section{Matrix Product Operator Entanglement}
\label{AppLogEnt}
In the MPO form of density matrix $\rho(\beta \equiv 1/T)$,
we compute the bipartite entanglement entropy
$S_E(\beta) = -\rm{Tr} (\tilde{\rho}_A  \ln \tilde{\rho}_A)$,
where $\tilde{\rho}_A$ is the reduced matrix of
subsystem $A$ (chosen as half the system)
in the superstate $| \rho(\beta) \rangle$ purified from the thermal density matrix $\rho(\beta)$.
In XTRG, $S_E(\beta)$ can be used to measure the required
computational resource, since the bond dimension scales as
$D \sim e^{S_E(\beta)}$ for an accurate simulation down to temperature $T \equiv 1/\beta$.
For a gapped system, $S_E$ typically saturated as $T$ is below
the excitation gap $\Delta$, and thus one can in principle simulate the
low-$T$ properties even down to the ground state with
a sufficiently large, but finite, bond dimension $D$.
The logarithmic entanglement scaling $S_E(\beta) \sim \eta \ln{\beta}$
requires $D \propto \beta^\eta$ with $\eta$
an algebraic exponent. The logarithmic entanglement
scaling at low temperature usually corresponds to gapless excitations.
Some typical examples include the low-$T$
thermal states near the 1+1D conformal critical point
and in 2D Heisenberg model with Goldstone modes \cite{Chen2018,Lih2019}.
In the former case, the exponent has been found to be universal,
$\eta =  c/3$, with $c$ the central charge of
the conformal field theory \cite{Chen2018}. 

In \Fig{FigEnt}(a), we show the bipartite entanglement entropies
$S_E$ across the center of the system under various magnetic fields
$h$, which exhibits a logarithmic scaling as
$S_E \sim \frac{2}{3}  \ln{\beta}$ in the intermediate-$T$ regime,
as indicated by the black dashed line.
In \Fig{FigEnt}(b,c), we have also computed the MPO entanglement
for the Kitaev-Heisenberg and Kitaev-Gamma models, where
similar logarithmic behaviors are also observed in the fractional liquid regime.
As the perturbative strength increases, for all three cases
in \Fig{FigEnt}(a-c), the entanglement $S_E$
deviates the logarithmic behavior, exhibits a
shoulder/peak and eventually converges to a constant
at sufficiently low temperature.

The exponent $\eta \approx 2/3$ suggests that the required
computational resource for simulating the Kitaev models
at intermediate temperature is similar to a conformal
critical point with central charge $c=2$.
Besides, in \Fig{FigEnt}(a) the maximal values of $S_E$
appears at the lowest temperature,
and is less than 3.5, which reveal the accessibility of
the YC$4\times6\times2$ systems by retaining not too many bond states.
Matter of fact, we have plotted in \Fig{FigEnt}(a) the $D=400$ and $600$
results, where a very good convergence can be seen.

\section{Magnetic susceptibility along various field directions}
\label{AppFieldOrients}

In \Fig{FigVarFields}, we show the $\chi$ data along various field directions,
as well as the case $h$ // [0 0 1] considered in the main text.
We label the field direction $l \, \hat{h}_x + m \, \hat{h}_y + n \, \hat{h}_z$ by [$l, m, n$],
where $\hat{h}_x, \hat{h}_y, \hat{h}_z$ are three cubic axes
in the spin space, therefore the Zeeman term reads
\begin{equation}
H_{f} = - \sum_i \frac{h}{\sqrt{l^2+m^2+n^2}} \, (l \, S_i^x + m \, S_i^y + n \, S_i^z ).
\end{equation}

Firstly, we apply an in-plane field $h$ // [-1 0 1] whose direction is shown
in \Fig{FigVarFields}(a), and compute the susceptibility. The results are shown in
\Fig{FigVarFields}(b,c), where an intermediate Curie behavior
can be clearly identified.
For the out-of-plane fields $h$ perpendicular to the \textit{ab} plane,
i.e., h // [1 1 1], there also emerges a Curie susceptibility [see \Fig{FigBnchmk}(f)],
as already discussed in Appendix~\ref{App:XTRG}.
As for more fields directions, we perform ED calculations on systems of very limited size, $N=12$ cluster,
and show the results in \Fig{FigVarFields}(d).
A diverging susceptibility behavior near the high-$T$ scale $T_H$ is also evident,
confirming that the emergent Curie-law susceptibility is robust and ubiquitous in the Kitaev model.
This universal susceptibility scaling can be observed
under various fields of different directions, on different lattice sizes/geometries,
and by different methods including ED, QMC, and XTRG approaches.

\begin{figure}[t!]
\includegraphics[angle=0,width=0.9\linewidth]{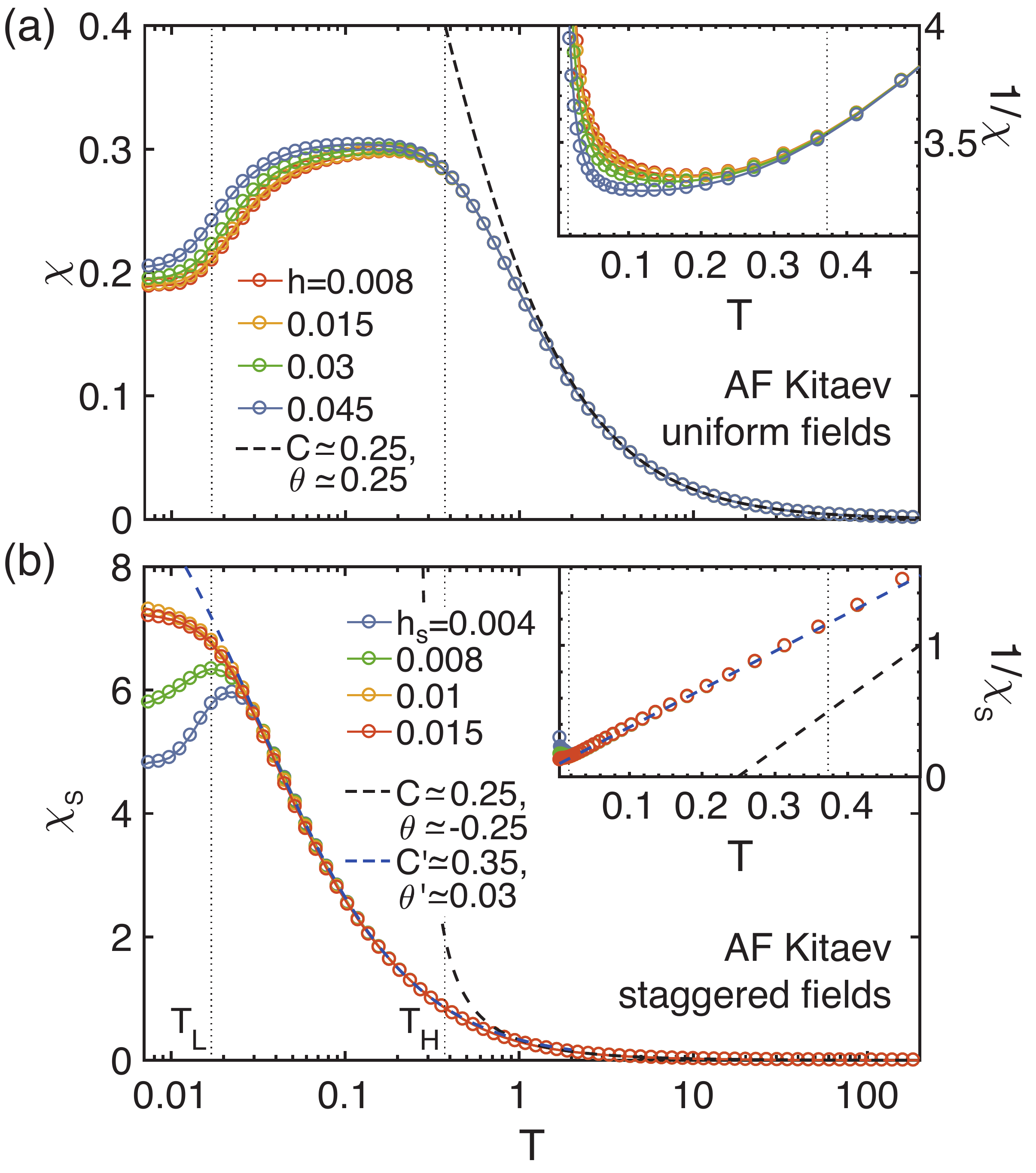}
\renewcommand{\figurename}{\textbf{Fig.}}
\caption{(a) Uniform susceptibility $\chi$ of the AF Kitaev model under
various fields $h$. 
The black dashed line indicates the high-$T$ Curie-Weiss fitting, and
at intermediate temperature no emergent Curie behavior is found,
which is also evident as shown in the $1/\chi$ plot (inset).
Since the AF and FM Kitaev models are exactly equivalent under zero fields,
here the two temperature scales $T_L$ and $T_H$ coincide with
those in \Fig{FigSus}(a) of the main text.
(b) The staggered susceptibility $\chi_S$ of AF Kitaev model
exhibits a universal Curie behavior at intermediate-$T$ regime (see also $1/\chi_S$ in the inset).
In fact, the $\chi_S$ data of AF Kitaev model bears a quantitative agreement
with the uniform susceptibility $\chi$ of the FM Kitaev model.
}
\label{FigAFM}
\end{figure}

\section{Kubo Formula of Magnetic Susceptibility}
\label{AppSusKitaev}

Here we derive the Kubo formula of the magnetic susceptibility,
in terms of imaginary-time dynamical correlations.
Suppose the magnetization of the system is uniform,
the magnetic moment per spin, under a perturbative magnetic field
$h$ along $z$-axis, is the expectation value of $S^z$ on arbitrary site $i_0$
\begin{equation}
  M = \langle S_{i_0}^z \rangle_{\beta} = \frac{1}{Z(\beta)} \operatorname{Tr} [ S_{i_0}^z e^{-\beta H} ]
\label{EqMag}
\end{equation}
where $H(K,h) = H_0(K) - hS^z_{\text{tot}}$, $H_0(K)$
is the pure Kitaev Hamiltonian, and $S^z_{\text{tot}} = \sum_j S^z_j$ is the total magnetization operator.

The magnetic susceptibility $\chi = \partial M /\partial h$ can be obtained by taking derivative of \Eq{EqMag}.
Note that since $[H, S^z_{\text{tot}}] \neq 0$, the derivative of the exponent
$-\beta \partial H /\partial h = -\beta S^z_{\text{tot}}$ can not be taken down straightforwardly
from the exponential, i.e.
\begin{equation}
  \frac{\partial}{\partial h} e^{-\beta H} \neq -\beta \frac{ \partial H}{\partial h} e^{-\beta H}.
\end{equation}
Instead, for an operator $O(u)$ as a function of $u$,
\begin{equation}
  \frac{\partial}{\partial u} e^{O} = \int_0^1 e^{\lambda O} \frac{ \partial O}{\partial u}  e^{(1-\lambda) O} d\lambda .
\label{EqDerivative}
\end{equation}
This can be proved via the decomposition $e^{O} = (e^{O/N})^N$, and for sufficiently large $N$,
it satisfies $\partial e^{O/N} / \partial u = e^{O/N} (\partial O/\partial u)/N$. Therefore,
\begin{equation}
\begin{aligned}
  \frac{\partial}{\partial u} e^{O} & = \sum_{n=1}^N (e^{O/N})^{(n-1)} \frac{\partial e^{O/N}}{\partial u} (e^{O/N})^{(N-n)} \\
  & \simeq \frac{1}{N}\sum_{n=1}^N e^{\frac{n}{N}O} \frac{ \partial O}{\partial u} e^{(1-\frac{n}{N})O}.
\end{aligned}
\end{equation}
Taking the limit $N\rightarrow \infty$, we arrive at \Eq{EqDerivative}.
With this, we can express the susceptibility as
\begin{equation}
\begin{aligned}
  \chi & = \frac{\beta}{Z(\beta)} \int_0^1 \operatorname{Tr} [ S_{i_0}^z e^{-\lambda\beta H} S^z_{\text{tot}} e^{-(1-\lambda)\beta H} ] d\lambda - \beta \langle S_{i_0}^z \rangle_\beta \langle S_{\text{tot}} ^z \rangle_\beta \\
  & = \frac{1}{Z(\beta)} \int_0^\beta \operatorname{Tr} [ S_{i_0}^z e^{-\tau H} S^z_{\text{tot}} e^{(\tau - \beta)H} ] d\tau -  \beta  \langle S_{i_0}^z \rangle_\beta \langle S_{\text{tot}} ^z \rangle_\beta \\
  & = \frac{1}{Z(\beta)} \int_0^\beta \operatorname{Tr} [ e^{\tau H} S_{i_0}^z e^{-\tau H} S^z_{\text{tot}} e^{-\beta H} ] d\tau -  \beta  \langle S_{i_0}^z \rangle_\beta \langle S_{\text{tot}} ^z \rangle_\beta \\
  & =  \sum_j \int_0^\beta \langle S_{i_0}^z (\tau) \, S_j^z \rangle_\beta \, d\tau -  \beta  \langle S_{i_0}^z \rangle_\beta \langle S_{\text{tot}} ^z \rangle_\beta,
\end{aligned}
\end{equation}
where on the second line we have made the variable substitution $\tau \equiv \lambda \beta$,
and in the last second line we perform a cyclic permutation within the trace.
For the sake of simplicity (and without loss of generality), we assume
$M=\langle S_{i_0}^z \rangle_\beta = 0$
in the discussion of fractional liquid, and thus arrive at the Kubo formula
\Eq{EqDynaSus} in the main text.

If the total spin $S^z_{\text{tot}}$ commutes with the Hamiltonian $H$,
e.g. in the Heisenberg model, the two factors $e^{\tau H}$ and $e^{-\tau H}$
cancel each other and the Kubo formula reduces to the conventional
susceptibility expression in terms of equal-time spin correlators.
Here in the Kitaev model, although $S^z_{\text{tot}}$ is not a good
quantum number, $e^{\tau H}$ and $e^{-\tau H}$ nevertheless
approximately cancel  in the fractional liquid regime
(see discussions in \Sec{SecSus}),
and \Eq{EqStatSus} of the main text also holds, which in turn gives
rise to a Curie behavior with a modified Curie constant $C'>1/4$.
However, as temperature further decreases to the KSL regime,
 ``dynamical" correlations decays exponentially as the gauge fluxes
freeze at low temperature below the finite flux excitation gap, which can
suppress the universal Curie-law susceptibility at intermediate temperature.

\section{Uniform and Staggered Magnetic Susceptibilities in the Antiferromagnetic Kitaev Model}
\label{AppKGAFSus}
For the isotropic Kitaev model with the AF Kitaev interaction
$K_{x,y,z} = 1$, we also compute the uniform magnetic susceptibility $\chi$ by
applying small fields that coupled to $S^z$ component, i.e.,  along the [0 0 1] direction.
In sharp contrast to the FM Kitaev model results in \Fig{FigSus} of the main text,
the uniform susceptibilities $\chi$ in \Fig{FigAFM}(a) do not exhibit a universal
Curie behavior in the intermediate-$T$ regime. The susceptibility curves there
are non-diverging convex curves and can not be fitted by the Curie-Weiss formula.
Our cylindrical results of AF Kitaev susceptibility in \Fig{FigAFM}(a)
are again in excellent agreement with previous QMC results \cite{Yoshitake2017b}.

The underlying reason for the absence of emergent Curie behaviors
under uniform fields is that the nearest spin pairs in the AF Kitaev model are
also correlated along $d_z$ bond, while in an anti-parallel manner.
Therefore, the change of total spin that relates to uniform susceptibility,
corresponds to flipping one of the two spins, which has to overcome a finite
amount of AF binding energy along the $d_z$ bond and does not reflect the
fluctuating fluxes. Matter of fact, as shown in \Fig{FigAFM}(b),
the staggered susceptibility $\chi_S$ under a staggered magnetic field $h_S$
(which has alternating sign on A and B sublattices of the Kitaev honeycomb)
exhibits a universal Curie behavior at intermediate temperature.

\begin{figure}[t!]
\includegraphics[angle=0,width=1\linewidth]{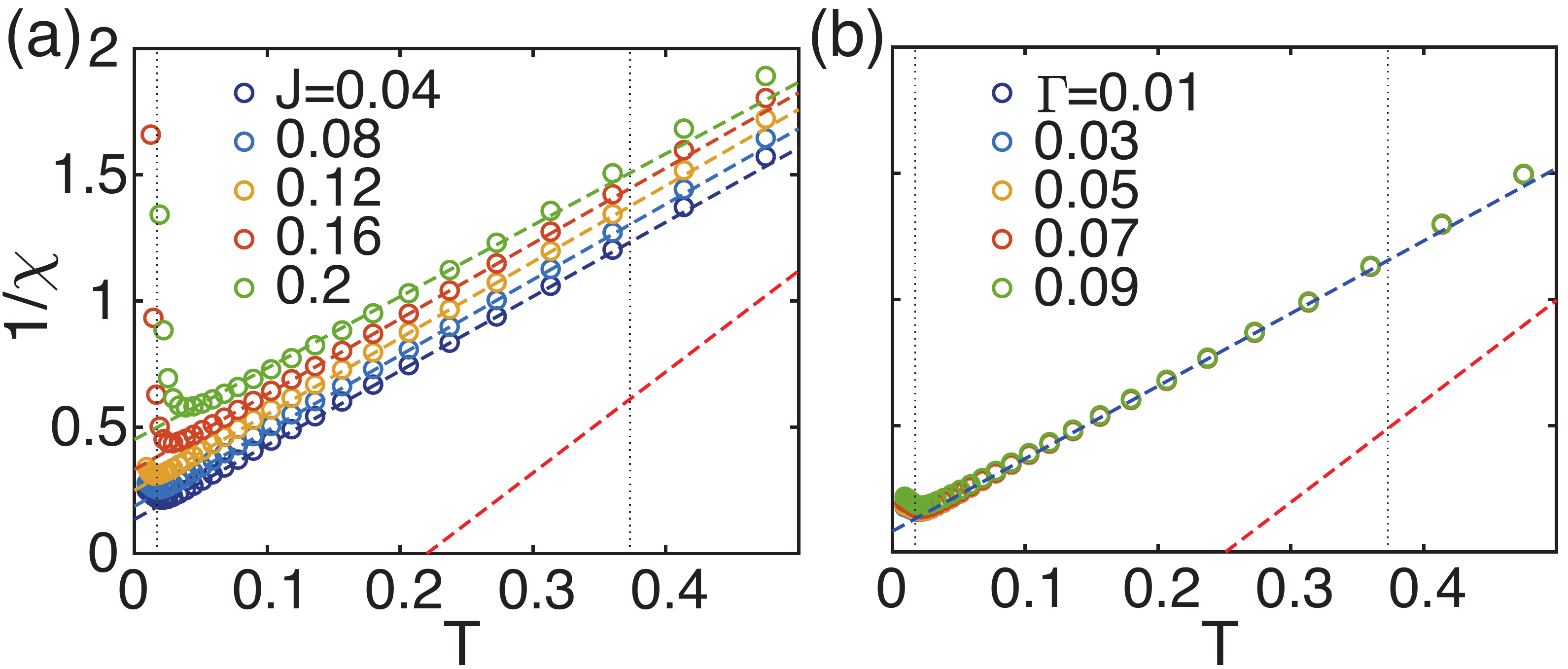}
\renewcommand{\figurename}{\textbf{Fig.}}
\caption{The inverse susceptibility $1/\chi$ of (a) the Kitaev-Heisenberg and
(b) Kitaev-Gamma models 
with various $J$ or $\Gamma$ perturbations.
The linear fittings are shown in both the high-$T$ (the red dashed lines) and intermediate-$T$ regimes,
with the same fitting parameters as those in \Fig{FigKHKG}(g, h)
of the main text.
}
\label{Fig1overChi}
\end{figure}

\begin{figure}[t!]
\includegraphics[angle=0,width=1\linewidth]{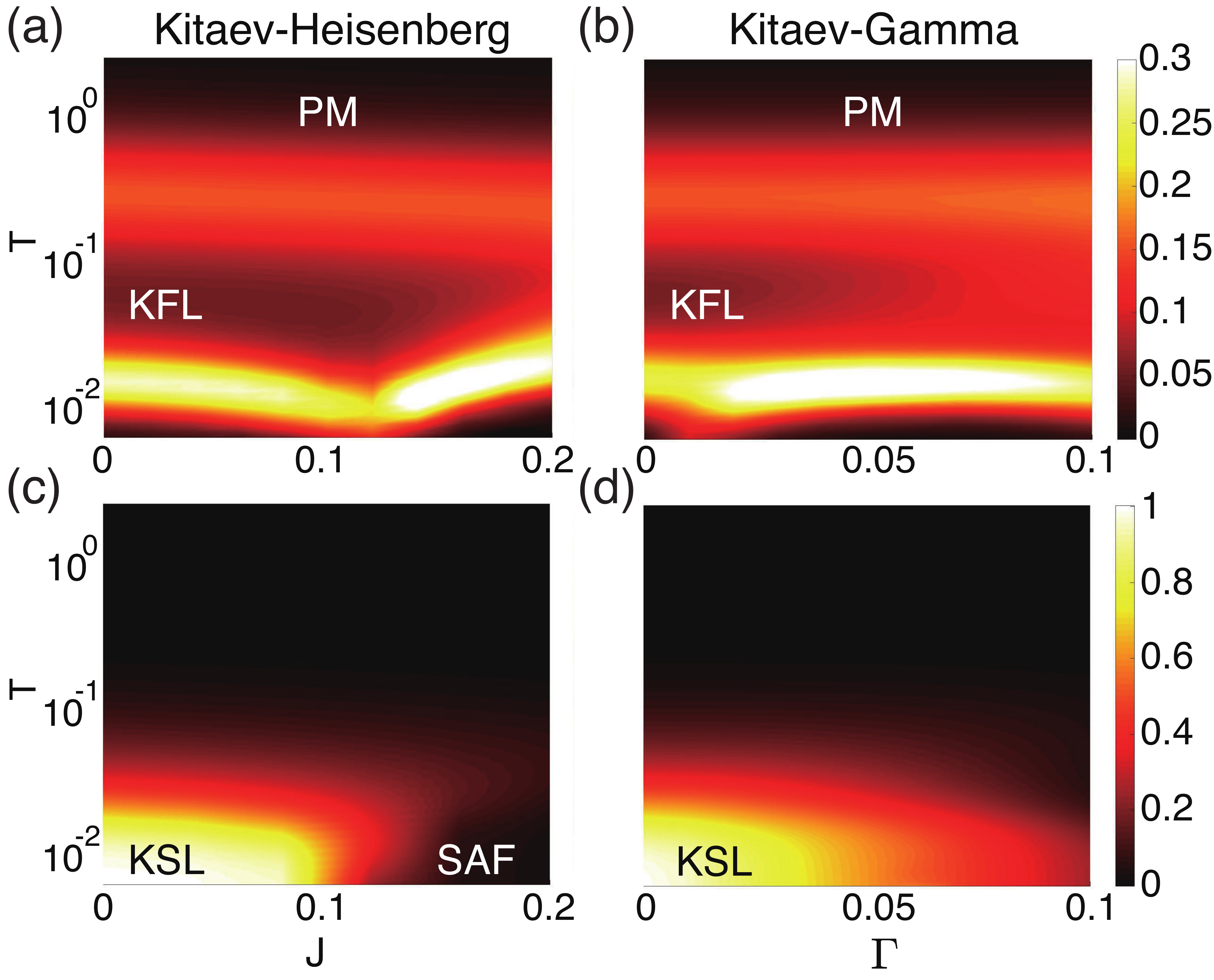}
\renewcommand{\figurename}{\textbf{Fig.}}
\caption{The landscape of specific heat data for the
(a) Kitaev-Heisenberg and (b) Kitaev-Gamma models. The corresponding
gauge flux $W_{\rm P}$ are contour plotted in (c) and (d) for the two models.
}
\label{FigContourKHKG}
\end{figure}

\section{Thermodynamics of the Kitaev-Heisenberg and Kitaev-Gamma Models}
\label{AppThermoKHKG}

In this section, we provide additional thermodynamic results of the Kitaev-Heisenberg 
and Kitaev-Gamma models. In \Fig{Fig1overChi}, we plot the inverse susceptibility data $1/\chi$ vs. $T$,
under the Heisenberg [\Fig{Fig1overChi}(a)] and Gamma [\Fig{Fig1overChi}(b)] perturbations.
The emergent Curie-Weiss susceptibility scaling is evident at intermediate temperature
in \Fig{Fig1overChi}, complementing the $\chi$ vs. $T$ fittings in \Fig{FigKHKG}(g,h) of the main text.

In \Fig{FigContourKHKG}, we show contour plots of the specific heat $C_V$
and gauge flux $W_{\rm P}$ results of both models. From \Fig{FigContourKHKG}(a,b),
we see that the $C_V$ curves are always double-peaked, which defines two
temperature scales $T_H$ and $T_L$ that bound the fractional liquid regime.
The corresponding flux expectations $W_{\rm P}$ are plotted in \Fig{FigContourKHKG}(c,d),
where the asymptotic KSL regimes with nonzero $W_{\rm P}$ can be clearly identified.

For the Kitaev-Heisenberg model, the low-$T$ scale $T_L$ determined from the lower peak 
position of $C_V$ exhibits a clear dip at $J_c \simeq 0.12$ in \Fig{FigContourKHKG}(a),
and for $J\lesssim J_c$ $W_{\rm P}$ quickly establishes itself at around $T_L$ as shown in \Fig{FigContourKHKG}(c).
The scenario is altered for $J\gtrsim J_c$, where different low-$T$ scale $T'_L$ emerges,
which represents the temperature scale between the SAF phase and Kitaev fractional liquid.

On the contrary, for the Kitaev-Gamma model in \Fig{FigContourKHKG}(b), the low-$T$ scale
$T_L$ remains stable for various $\Gamma$ values, suggesting that the low-$T$
KSL regime is very robust against Gamma perturbations. This is also supported
by the flux results in \Fig{FigContourKHKG}(d), where $W_{\rm P}$ remains nonzero
at low temperature at least up to $\Gamma=0.1$.

\AtEndEnvironment{thebibliography}{
}
\bibliography{kitaevRef}

\end{document}